\documentclass{article}
\usepackage{amsmath,amsthm}
\usepackage{graphicx,epstopdf,epsfig,multirow,epic,bm}
\usepackage{lineno,hyperref}
\usepackage{subfigure}
\usepackage{amsfonts} 
\usepackage{algorithm}  
\usepackage{algpseudocode}   
\normalsize \rm
\begin{document}

\begin{center}
{\Large  \textbf { Type-II Apollonian Model }}\\[12pt]
{\large Fei Ma$^{a,}$\footnote{~The author's E-mail: mafei123987@163.com. }, Jinzhi Ouyang$^{a}$,  Ping Wang$^{b,c,d,}$\footnote{~The author's E-mail: pwang@pku.edu.cn.}, Haobin Shi$^{a}$, and Wei Pan$^{a}$}\\[6pt]
{\footnotesize $^{a}$ School of Computer Science, Northwestern Polytechnical University, Xi'an 710072, China\\
$^{b}$ National Engineering Research Center for Software Engineering, Peking University, Beijing, China\\
$^{c}$ School of Software and Microelectronics, Peking University, Beijing  102600, China\\
$^{d}$ Key Laboratory of High Confidence Software Technologies (PKU), Ministry of Education, Beijing, China}\\[12pt]
\end{center}

\begin{quote}
\textbf{Abstract:} The family of planar graphs is a particularly important family and models many real-world networks. In this paper, we propose a principled framework based on the widely-known Apollonian packing process to generate new planar network, i.e., Type-II Apollonian network $\mathcal{A}_{t}$. The manipulation is different from that of the typical Apollonian network, and is proceeded in terms of the iterative addition of triangle instead of vertex. As a consequence, network $\mathcal{A}_{t}$ turns out to be hamiltonian and eulerian, however, the typical Apollonian network is not. Then, we in-depth study some fundamental structural properties on network $\mathcal{A}_{t}$, and verify that network $\mathcal{A}_{t}$ is sparse like most real-world networks, has scale-free feature and small-world property, and exhibits disassortative mixing structure. Next, we design an effective algorithm for solving the problem of how to enumerate spanning trees on network $\mathcal{A}_{t}$, and derive the asymptotic solution of the spanning tree entropy, which suggests that Type-II Apollonian network is more reliable to a random removal of edges than the typical Apollonian network. Additionally, we study trapping problem on network $\mathcal{A}_{t}$, and use average trapping time as metric to show that Type-II Apollonian network $\mathcal{A}_{t}$ has better structure for fast information diffusion than the typical Apollonian network. \\


\end{quote}

\vskip 1cm

\section{Introduction}

In the past two decades, complex network, as a theoretical tool for understanding complex systems, has been proven powerful and useful in a wide variety of practical applications in different fields ranging from statistical physics, applied mathematics to computer science, even to chemistry, and so on \cite{Barab-2016}-\cite{Tyloo-2018}. Consequently, some ubiquitous properties hid in real-world complex systems have been uncovered, including scale-free feature \cite{Albert-1999-1}, small-world effect  \cite{Watts-1998}, and self-similarity \cite{Song-2005} etc.   

In the study of complex networks, it is of great interest to establish theoretical models that reliably display some popularly-observed properties mentioned above. The reason for this is particularly because doing so is helpful to investigate the evolution mechanism of complex systems. A great number of theoretical models have been proposed. Consider for instance the small-world effect that indicates that most real-life networks have a small diameter and a higher clustering coefficient \cite{Amaral-2000,Latora-2001}. One of most well-known networked models for depicting such a phenomenon is the WS-model attributed to Watts and Strogatz \cite{Watts-1998}. Again for instance, the scale-free feature suggests that power-law degree distribution is observed on a large number of networks \cite{Barab-2016}. The BA-model built by Barab\'{a}si and Albert \cite{Albert-1999-1} is one of most famous theoretical models for illustrating scale-free phenomena. Among them, a lot of networked models are established based on some classic graphs such as Apollonian graph \cite{Andrade-2005,Doye-2005} , Sierpinski gasket \cite{Chang-2007,Ri-2020}, Vicsek fractal \cite{Vicsek-1991,Ma-epl-2021}, T-graph \cite{Agliari-2008,Ma-tkde-2020}, dense graph\cite{Ma-2020-T-1,Ma-pre-2020}, cellular neural networks \cite{Liu-2020} and so on. These models have been studied in more detail. Consequently, the majority of popular properties mentioned above can be observed on these models. What's more, most of them are planar graphs. The family of planar graphs is a particularly important family and models important real-world graphs such as road networks, the layout of printed circuits,
river networks upon the earth's surface, and so forth \cite{Mahapatra-2021,Dujmovic-2021}. Motivated by this, in this work, we propose a framework to generate a class of networked models, and show that the proposed model indeed has those widely-observed properties.

In addition, the study of various dynamics taking place on networks has also attracted more attention from a wide range of fields, and has been becoming a hot-topic in the realm of complex networks \cite{Newman-2011}. Among which, trapping problem is an integral major theme of dynamics \cite{Masuda-2017}. In fact, this is a specific case of random walks (diffusion) where a trap is positioned at a given location on network, which absorbs all particles visiting it. In general, one chooses the location with respect to what he/she is interested in, such as, the greatest degree vertex is often selected to serve as location.  Trapping problem is closely relevant to a wide range of applications and has led to a great number of theoretical and practical investigations in the past decades \cite{Yen-2010,Condamin-2008,Huang-2021}. A fundamental question in the study of trapping problem on network is to determine a quantity called the average trapping time. This important parameter has been proven helpful in the study of transport-limited reactions \cite{Loverdo-2008}, target search \cite{Zhao-2022,Zaheer-2022}, and so on.

It should be mentioned that we build up theoretical model based on Apollonian packing \cite{Kasner-1943}. In the widely-known packing process, which dates back to Apollonius of Perga (c262-c190BC), one begins with three mutually tangent circles, and then introduces a new circle into the region enclosed by three curves each of which is selected from a circle. At the same time, the newly added circle is required to be tangent to the three circles. Next he (or she) adds a circle into each inner region enclosed by this circle and each pair of the original three. Such a process is iterated, see Fig.1 for more detailed information. One thing to note is that such a packing process has been used to create a class of networked models \cite{Andrade-2005,Doye-2005}, called the typical Apollonian model for our purpose and convenience. The model introduced below, however, differs from the pre-previous models, which is elaborated in the rest of this paper. For brevity, throughout this paper, the terms graph and network are used interchangeably and denoted by $\mathcal{G}=(\mathcal{V},\mathcal{E})$. At the same time, $|\mathcal{V}|$ is the number of vertices and $|\mathcal{E}|$ is the number of edges. Symbol $[a,b]$ represents an ensemble of integers $\{a, a+1,...,b\}$. 

The main contribution of this work is as follows.

(1) We propose a new network based on the widely-known Apollonian packing process, which is called Type-II Apollonian network $\mathcal{A}_{t}$. This network turns out to be both hamiltonian and eulerian. In the meantime, network $\mathcal{A}_{t}$ is a maximal planar graph. 

(2) We study some structural properties of Type-II Apollonian network $\mathcal{A}_{t}$, and show that scale-free feature and small-world property can be observed on this network. In addition, network $\mathcal{A}_{t}$ is proved to be disassortative. 

(3) We design an efficient algorithm for counting the total number of spanning trees of network $\mathcal{A}_{t}$, and then prove that Type-II Apollonian network $\mathcal{A}_{t}$ is more reliable to a random removal of edges than the typical Apollonian network. 

(4) We discuss trapping problem on Type-II Apollonian network $\mathcal{A}_{t}$, and derive the asymptotic solution of average trapping time. Based on this, Type-II Apollonian network $\mathcal{A}_{t}$ is proved to have better structure for fast information diffusion than the typical Apollonian network. 

\begin{figure}
\centering
\includegraphics[height=7cm]{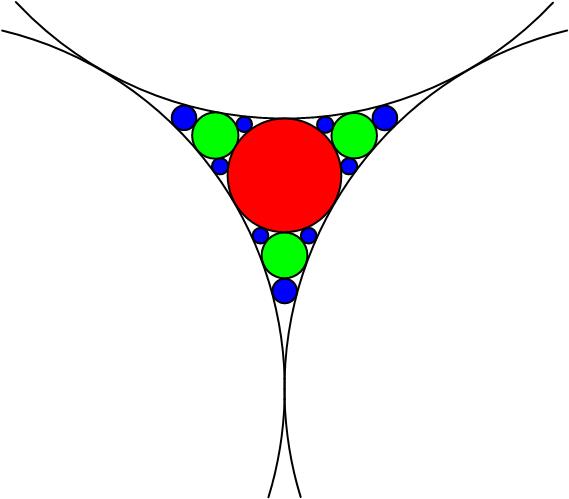}%\includegraphics[width=3in]{model2_ExpIncomeperTime_SymmetricVsAsymmetric.eps}
\caption{(Color online) The first steps of the well-known Apollonian packing process.}\label{fig:ExpIncomePerTime}
\end{figure}

\emph{Roadmap}---The rest of this paper is organized as follows. Section 2 introduces a generative framework for producing a new network model, called Type-II Apollonian network $\mathcal{A}_{t}$, and then calculates vertex number and edge number of network $\mathcal{A}_{t}$. Next, in Section 3, some fundamental parameters associated with topological structure of network $\mathcal{A}_{t}$ are discussed in more detail, including average degree, degree distribution, clustering coefficient, diameter and Pearson correlation coefficient. Following the above section, we consider the problem of how to enumerate spanning trees on network $\mathcal{A}_{t}$ in Section 4. As a consequent, we propose an effective algorithm for counting spanning trees, and also derive the corresponding spanning tree entropy. Section 5 is concerned with trapping problem on network  $\mathcal{A}_{t}$. Particularly, we derive the exact or asymptotic solution of average trapping time in two distinct settings. Related work is briefly reviewed in Section 6. Finally, we close this paper in Section 7.

\section{Type-II Apollonian network $\mathcal{A}_{t}$}

In this section, we will propose a framework for constructing Type-II Apollonian network $\mathcal{A}_{t}$. It is worth noticing that at first sight, the resulting network seems not closely related to the well-known Apollonian packing. Yet, we will provide a detailed explanation about how to establish a close connection between them. In the meantime, we also explain why the proposed network is called Type-II Apollonian network.

\textbf{\emph{Framework}}
\begin{itemize}

\item At $t=1$, the seminal model, denoted by $\mathcal{A}_{1}$, is a triangle as shown in the left-most panel of Fig.2.

\item  At $t=2$, we add three active vertices into seed $\mathcal{A}_{1}$ each of which is assigned to an edge of model $\mathcal{A}_{1}$, and then connect each active vertex to two end-vertices of its corresponding edge. In addition, these active vertices are connected by three new edges so as to generate a triangle. After that, the resulting network is $\mathcal{A}_{2}$. An illustrative example is plotted in the middle panel of Fig.2 where these newly added active vertices are highlighted (indigo online).

\item At $t=3$, for each triangle in network $\mathcal{A}_{2}$ that only contains an active vertex, we execute the same operation as above. Meanwhile, those vertices added at time step $2$ will become inactive. The resultant graph is network $\mathcal{A}_{3}$ as shown in the right-most panel of Fig.2.

\item At $t>3$, the newborn model $\mathcal{A}_{t}$ is constructed from the preceding model $\mathcal{A}_{t-1}$ in a similar manner mentioned above.

     \emph{Remark} In essence, we contain a new triangle with three active vertices into each prescribed triangle at each time step $t(\geq2)$. Here, a triangle is selected as prescribed one if it only contains an active vertex. For our purpose, an edge connecting two active vertices is also considered active. Also, an edge is considered semi-active if one end-vertex is active yet the other is not.

\end{itemize}

\begin{figure}
\centering
\includegraphics[height=4cm]{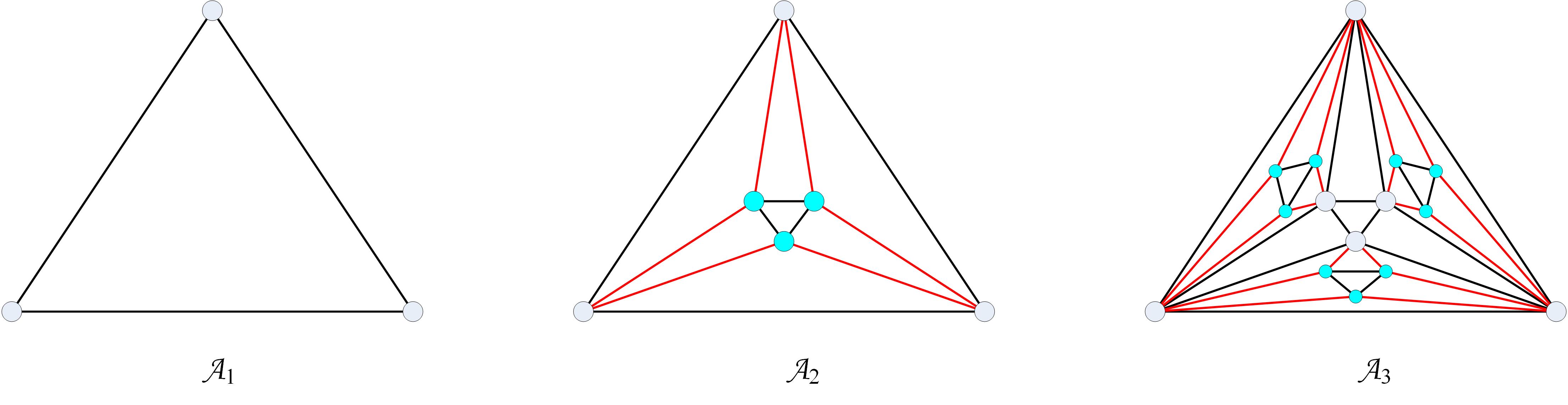}%\includegraphics[width=3in]{model2_ExpIncomeperTime_SymmetricVsAsymmetric.eps}
\caption{(Color online) The diagram of the first three generations of Type-II Apollonian network $\mathcal{A}_{t}$.}\label{fig:ExpIncomePerTime}
\end{figure}

After $t$ time steps, we obtain the desirable network $\mathcal{A}_{t}$. Due to this generative manner, two fundamental structural parameters of model $\mathcal{A}_{t}$, i.e., vertex number $|\mathcal{V}_{t}|$ and edge number $|\mathcal{E}_{t}|$, are immediately obtained. For smaller value of $t$, the concrete values of $|\mathcal{V}_{t}|$ and $|\mathcal{E}_{t}|$ can be obtained by hand, such as $|\mathcal{V}_{1}|=3$, $|\mathcal{E}_{1}|=3$; $|\mathcal{V}_{2}|=6$, $|\mathcal{E}_{2}|=12$. For larger $t(\geq3)$, it is also not hard to write
\begin{equation}\label{eq:MF-2023-2-1}
|\mathcal{V}_{t}|=3|\mathcal{V}_{t-1}|-3,\qquad |\mathcal{E}_{t}|=3|\mathcal{E}_{t-1}|+3.
\end{equation}
With previous conditions, Eq.(\ref{eq:MF-2023-2-1}) can be solved to yield the exact solutions of quantities $|\mathcal{V}_{t}|$ and $|\mathcal{E}_{t}|$, as below:

 \begin{equation}\label{eq:MF-2023-2-2}
|\mathcal{V}_{t}|=\frac{1}{2}\times3^{t}+\frac{3}{2},\qquad |\mathcal{E}_{t}|=\frac{1}{2}\times3^{t+1}-\frac{3}{2}.
\end{equation}

Now, let us divert attention to establish relationship between network $\mathcal{A}_{t}$ and the well-known Apollonian packing. Based on the description about Apollonian packing in Introduction, these points of tangency in three mutually tangent circles are abstracted into three vertices, and then they are connected into a triangle (also called $K_{3}$ or $3$-cycle in the jargon of graph theory). Subsequently, each circle added into Apollonian packing is replaced with a triangle. In Apollonian packing, introduction of each circle leads to three points of tangency. Therefore, the total number of points of tangency in Apollonian packing is completely equal to vertex number of network $\mathcal{A}_{t}$. To make more concrete, an illustrative example is shown in Fig.3.

Next, we briefly recall the typical Apollonian network $A_{t}$, and distinguish the proposed network $\mathcal{A}_{t}$ and network $A_{t}$. Reader is encouraged to refer Ref.\cite{Andrade-2005,Doye-2005} for more information about network $A_{t}$. The typical Apollonian network $A_{t}$ is built via iteratively replacing each added circle with a vertex where a pair of vertices are connected if their corresponding circles are tangent. So, the total number of circles in Apollonian packing is completely equal to vertex number of network $A_{t}$. Vertex number $|V_{t}|$ and edge number $|E_{t}|$ of network $A_{t}$ satisfy $|V_{t}|=\frac{3^{t-1}+5}{2}$ and  $|E_{t}|=\frac{3^{t}+3}{2}$. This graph is also known as two-dimensional Apollonian network, and was introduced by Andrade et al. \cite{Andrade-2005} and independently proposed by Doye and Massen \cite{Doye-2005}. It is clear to the eye that the proposed network $\mathcal{A}_{t}$ has a completely different topological structure from the typical Apollonian network $A_{t}$. For convenience, the typical Apollonian network $A_{t}$ is regarded as Type-I Apollonian network hereafter. 

Based on the demonstration above, the proposed network $\mathcal{A}_{t}$ is viewed as Type-II Apollonian network. It is now easy to see that these two classes of Apollonian networks are in fact generated by virtue of distinct viewpoints although they are based on the well-known Apollonian packing. Besides that, network $\mathcal{A}_{t}$ has some other interesting structural features. Here we just list out two of them. By definition in \cite{Bondy-2008}, network $\mathcal{A}_{t}$ is in practice a \emph{maximal planar graph} and is also \emph{hamiltonian} because it contains a hamiltonian cycle. As shown in Fig.4, a hamiltonian cycle of network $\mathcal{A}_{3}$ is highlighted in color (blue online). More generally, one can without difficulty prove that our assertion holds on Type-II Apollonian network $\mathcal{A}_{t}$ for arbitrary time step $t$. Hence, network $\mathcal{A}_{t}$ can embed the longest linear array between any two distinct nodes with dilation, congestion, load, and expansion all equal to one \cite{Fu-2005}. On the other hand, it is obvious to see that Type-I Apollonian network is not hamiltonian because no a hamiltonian cycle can be found as $t>2$. Similarly, by definition in \cite{Bondy-2008}, network $\mathcal{A}_{t}$ turns out to be \emph{eulerian} but Type-I Apollonian network is not. In a nutshell, this suggests that Type-II Apollonian network is more notable to deeply discuss in graph theory and theoretical computer science, which is left as the future work.

\begin{figure}
\centering
\includegraphics[height=6cm]{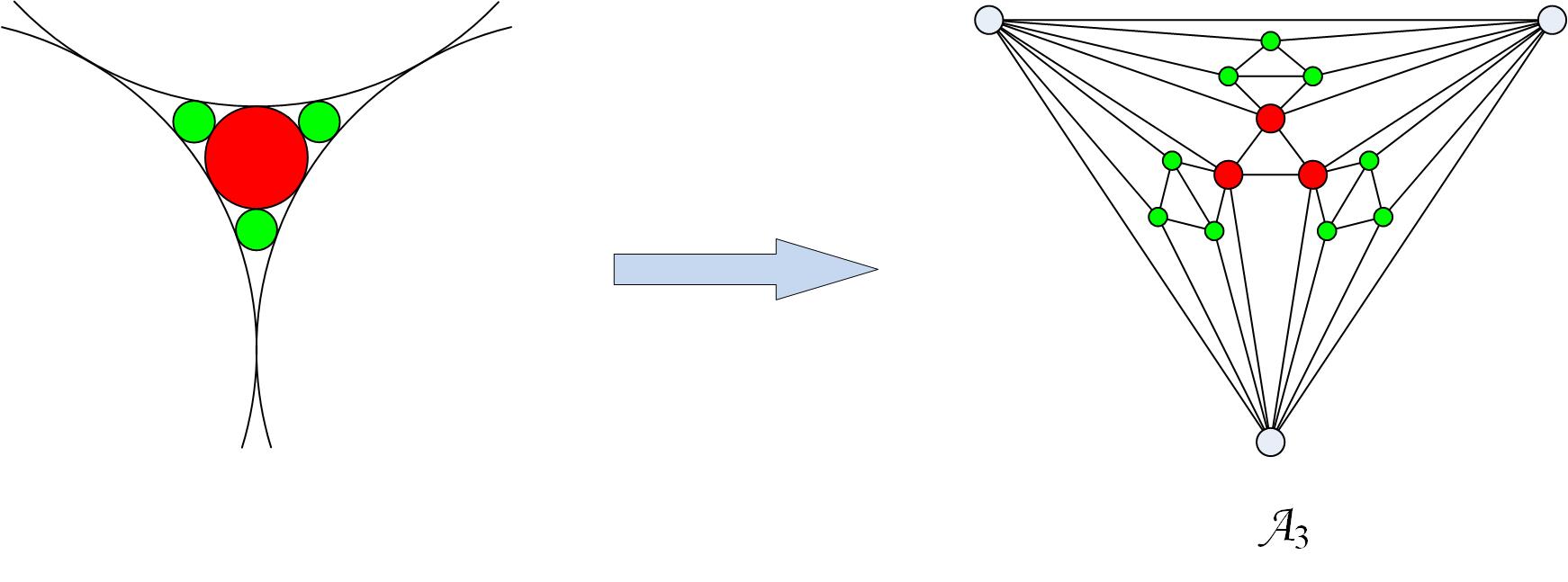}%\includegraphics[width=3in]{model2_ExpIncomeperTime_SymmetricVsAsymmetric.eps}
\caption{(Color online) An illustrative example for the transformation from the Apollonian packing process to Type-II Apollonian network $\mathcal{A}_{t}$. Here, we take case of $t=3$ as an example.}\label{fig:ExpIncomePerTime}
\end{figure}

The subsequent sections will focus mainly on discussing some other topological structure parameters of network $\mathcal{A}_{t}$, such as, average degree, degree distribution, clustering coefficient, diameter and Pearson correlation coefficient. At the same time, the problem of how to count spanning trees on network $\mathcal{A}_{t}$ is also considered in more detail. In addition, trapping problem on network $\mathcal{A}_{t}$ is deeply studied, and the solution of average trapping time is derived. During the next discussions, we will compare network $\mathcal{A}_{t}$ with network $A_{t}$ based on the obtained results in order to further clarify the significance of constructing and discussing the former.

\section{Structural properties}

In the context of complex networks \cite{Barab-2016,Newman-2018}, some structural features, such as, small-world property, are often measured by utilizing some structural parameters. Along such a research line, we now unmask some topological properties behind network $\mathcal{A}_{t}$ by determining the correspondingly structural parameters, including average degree, degree distribution, clustering coefficient, etc. Now, let us begin with estimating average degree of network $\mathcal{A}_{t}$ to determine whether it is sparse or not.

\subsection{Average degree}

As one of simplest yet most useful structural parameters, average degree is often chose as a measure to determine whether a given network is sparse or not \cite{Barab-2016}. In the language of mathematics, average degree $\langle k\rangle$ of a graph $\mathcal{G}=(\mathcal{V},\mathcal{E})$ is define as the ratio of the summation over degrees $k_{v}$ of all vertices to vertex number, i.e., $\langle k\rangle=\sum_{v\in \mathcal{V}}k_{v}/|\mathcal{V}|=2|\mathcal{E}|/|\mathcal{V}|$. In the literature, an evolving graph $\mathcal{G}=(\mathcal{V},\mathcal{E})$ is considered sparse if its average degree $\langle k\rangle$ is asymptotically equivalent to a constant $O(1)$ in the limit of large graph size, and dense otherwise. In the past twenty years, a large number of empirical observations have shown that the sparsity feature is ubiquitous in various complex networks \cite{Newman-2018}. Next, let us decide whether or not network $\mathcal{A}_{t}$ is sparse.

\textbf{Proposition 1} The average degree $\langle k_{t}\rangle$ of network $\mathcal{A}_{t}$ is

\begin{equation}\label{eq:MF-2023-3-1-0}
\langle k_{t}\rangle \sim6.
\end{equation}

\emph{Proof} By definition, the average degree $\langle k_{t}\rangle$ of network $\mathcal{A}_{t}$ is written as

\begin{equation}\label{eq:MF-2023-3-1-2}
\langle k_{t}\rangle=\frac{2|\mathcal{E}_{t}|}{|\emph{V}_{t}|}.
\end{equation}
After substituting the results from Eq.(\ref{eq:MF-2023-2-2}) in the above equation, we have

\begin{equation}\label{eq:MF-2023-3-1-3}
\langle k_{t}\rangle=\frac{2\times3^{t+1}-6}{3^{t}+3} \sim6.
\end{equation}
This reveals that network $\mathcal{A}_{t}$ is \emph{sparse} as some previously published networked models including Type-I Apollonian network \cite{Andrade-2005,Doye-2005}. \qed

\begin{figure}
\centering
\includegraphics[height=7cm]{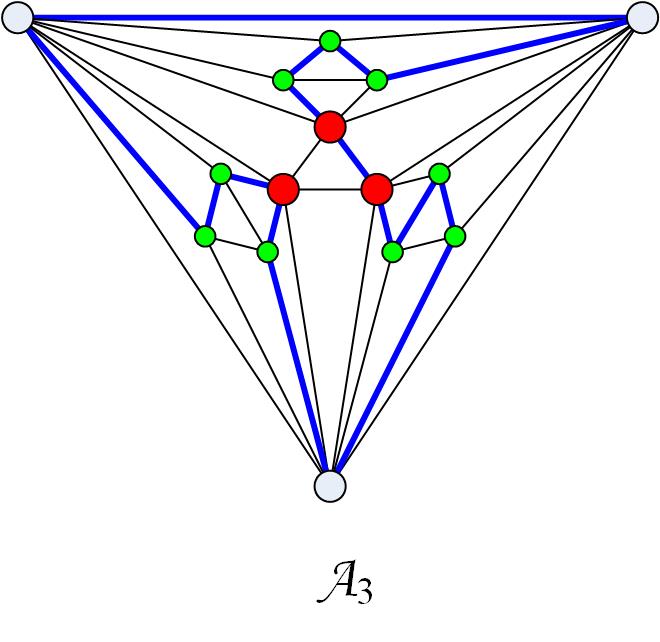}%\includegraphics[width=3in]{model2_ExpIncomeperTime_SymmetricVsAsymmetric.eps}
\caption{(Color online) The diagram of a hamiltonian cycle in network $\mathcal{A}_{3}$, which is highlighted in blue. }\label{fig:ExpIncomePerTime}
\end{figure}

\subsection{Degree distribution}

Degree distribution, which is closely associated with degree sequence, has better studied over the past years. While there are still some problems related to degree sequence unknown to one, such as, no better scheme for accurately determining whether a set of numbers is graphic or not \cite{Bondy-2008}, it is using degree distribution that some significant topological properties rooted on the underlying structure of complex networks have been unveiled. For instance, a great variety of real-world networks turn out to follow the highly right-skewed degree distribution \cite{Albert-1999-1}. Often, it is necessary for a given graph $\mathcal{G}=(\mathcal{V},\mathcal{E})$ to employ the cumulative degree distribution instead of degree distribution when quantifying the distribution rule of its vertex degrees. One main reason for this is that vertex degrees of graph $\mathcal{G}=(\mathcal{V},\mathcal{E})$ are discrete in form.

Therefore, in order to determine which type of degree distribution network $\mathcal{A}_{t}$ follows, we first need to introduce the definition of cumulative degree distribution \cite{Barab-2016}. For a given graph $\mathcal{G}=(\mathcal{V},\mathcal{E})$, its cumulative degree distribution $P_{cum}(k)$ is given by

\begin{equation}\label{eq:MF-2023-3-2-1}
P_{cum}(k)=\frac{\sum_{k_{i}\geq k}\mathcal{N}_{k_{i}}}{|\mathcal{V}|}
\end{equation}
here $\mathcal{N}_{k_{i}}$ is the total number of vertices with degree $k_{i}$ in graph $\mathcal{G}=(\mathcal{V},\mathcal{E})$.

Now, we study the degree distribution of network $\mathcal{A}_{t}$. As explained in Eq.(\ref{eq:MF-2023-3-2-1}), the first step is to classify vertices in the proposed network $\mathcal{A}_{t}$ in terms of vertex degree. With the help of detailed demonstration in Framework, it is clear to see that more earlier vertex is added, more larger the associated degree is. The following is a concrete derivation.

\textbf{Proposition 2} The cumulative degree distribution of network $\mathcal{A}_{t}$ obeys

\begin{equation}\label{eq:MF-2023-3-2-2}
P_{cum}(k_{t;i})\sim k_{t;i}^{-\gamma}, \quad \gamma=\frac{\ln3}{\ln2}.
\end{equation}

\emph{Proof} In network $\mathcal{A}_{t}$, the greatest degree vertices are those ones added at time step $1$ and their degrees $k_{t;1}$ are equal to $2^{t}$. The second greatest degree vertices are three ones introduced at time step $2$ and they all have degree (denoted by $k_{t;2}$) equal to $2^{t-1}+2$. Analogously, one has an ability to obtain that the total number of vertices with degree $k_{t;3}=2^{t-2}+2$, $\dots$, $k_{t;i}=2^{t-i+1}+2$, $\dots$, $k_{t;t}=4$, is equivalent to $\mathcal{N}_{t;3}=9$, $\dots$, $\mathcal{N}_{t;i}=3^{i-1}$, $\dots$, $\mathcal{N}_{t;t}=3^{t-1}$, respectively. It is worth noting that other values not mentioned here are absent in degree sequence of network $\mathcal{A}_{t}$. Then, based on Eq.(\ref{eq:MF-2023-3-2-1}), we write

\begin{equation}\label{eq:MF-2023-3-2-3}
P_{cum}(k_{t;i})=\frac{\sum_{j=1}^{i}\mathcal{N}_{t;j}}{\frac{1}{2}\times3^{t}+\frac{3}{2}}\sim 3^{i-t}.
\end{equation}

As mentioned above, the degree value $k_{t;i}$ is equal to $2^{t-i+1}+2$. Plugging $t-i\sim \ln\frac{k_{t;i}}{2}$ into the above equation yields

\begin{equation}\label{eq:MF-2023-3-2-4}
P_{cum}(k_{t;i})\sim k_{t;i}^{-\gamma}, \quad \gamma=\frac{\ln3}{\ln2},
\end{equation}
implying that network $\mathcal{A}_{t}$ follows power-law degree distribution with exponent $1+\frac{\ln3}{\ln2}$, and thus has \emph{scale-free feature}. \qed

One thing to note is that Type-I Apollonian network $A_{t}$ has also scale-free feature, and its power-law exponent is identical to $1+\frac{\ln3}{\ln2}$ as well. This suggests that such two types of Apollonian networks have the same degree distribution. On the other hand, they have different degree sequences, see \cite{Andrade-2005,Doye-2005} for more information about degree distribution of Type-I Apollonian network $A_{t}$.

\subsection{Clustering coefficient}

The cluster phenomena turn out to be popular on various real-world networks by means of estimation of a structural parameter, i.e., clustering coefficient \cite{Newman-2003}. For a vertex $v$ with degree $k_{v}$ in network $\mathcal{G}=(\mathcal{V},\mathcal{E})$, its clustering coefficient $c_{v}$ is formally defined in the following form

\begin{equation}\label{eq:MF-2023-3-3-1}
c_{v}=\frac{n_{v}}{\frac{k_{v}(k_{v}-1)}{2}},
\end{equation}
where $n_{v}$ is the number of edges actually existing between vertex $v$'s neighbors. For the whole network $\mathcal{G}=(\mathcal{V},\mathcal{E})$, the clustering coefficient is naturally defined by the averaged value over all vertices, i.e.,
\begin{equation}\label{eq:MF-2023-3-3-2}
\langle c\rangle=\frac{\sum_{v\in \mathcal{V}}c_{v}}{|\mathcal{V}|}.
\end{equation}
Often, a non-zero value of $\langle c\rangle$ means that there exists the so-called cluster phenomena on network under consideration. The larger $\langle c\rangle$ indicates the higher cluster phenomena.

\textbf{Proposition 3} The clustering coefficient $\langle c_{t}\rangle$ of network $\mathcal{A}_{t}$ is

\begin{equation}\label{eq:MF-2023-3-3-3}
\begin{aligned}\langle c_{t}\rangle&=\frac{2}{3^{t-1}+1}\times\frac{3\times2^{t-2}-1}{2^{t-2}(2^{t}-1)}\\
&\quad+\frac{2}{3^{t-1}+1}\times\sum_{i=1}^{t-1}3^{i-1}\times\frac{3\times2^{t-i-1}+1}{(2^{t-i-1}+1)(2^{t-i}+1)}.
\end{aligned}
\end{equation}

\emph{Proof} By definition, we need to calculate the exact solution of clustering coefficient $c_{v}$ of each vertex $v$ in network $\mathcal{A}_{t}$. Due to the analysis in subsection 3.2, it is obvious to see that an arbitrary vertex added at the same time has an identical degree. In addition, these vertices also have a clustering coefficient in common because the total number of edges actually existing between vertex's neighbors is identical. Therefore, let us determine the total number of edges that actually exist between each type of vertex's neighbors, as follows.

From the growth way of network $\mathcal{A}_{t}$, it is easy to see that at time step $j$, $3\times2^{j-1}$ new edges will emerge in the neighboring set of each vertex added at time step $1$. Hence, the total number $n_{t;1}$ of edges, which exist between neighbors of vertex added at time step $1$, is calculated to equal $4+\frac{3}{2}\sum_{j=1}^{t-1}2^{j}$. Analogously, for each vertex added at time $1<i\leq t$, the total edge number $n_{t;i}$ is easily derived to express $4+\frac{3}{2}\sum_{j=1}^{t-i}2^{j}$. We then write

\begin{equation}\label{eq:MF-2023-3-3-4}
\begin{aligned}\langle c_{t}\rangle&=\frac{1}{\frac{1}{2}\times3^{t}+\frac{3}{2}}\times3\times2\times\frac{4+\frac{3}{2}\sum_{j=1}^{t-1}2^{j}}{k_{t;1}\left(k_{t;1}-1\right)}\\
&+\frac{1}{\frac{1}{2}\times3^{t}+\frac{3}{2}}\times\sum_{i=2}^{t}3^{i-1}\times2\times\frac{4+\frac{3}{2}\sum_{j=1}^{t-i}2^{j}}{k_{t;i}\left(k_{t;i}-1\right)},
\end{aligned}
\end{equation}
in which we have used quantity $\mathcal{N}_{t;j}$. After some simple arithmetics, Eq.(\ref{eq:MF-2023-3-3-4}) is proved to be the completely same as Eq.(\ref{eq:MF-2023-3-3-3}) as desired.\qed

To make the analysis above more concrete, we provide an illustrative example where the clustering coefficients of the proposed network $\mathcal{A}_{t}$ are plotted into Fig.5, which suggests that network $\mathcal{A}_{t}$ has a higher clustering coefficient. Intuitively, it is particularly because there are enough triangles added in network $\mathcal{A}_{t}$. It should be mentioned that in \cite{Andrade-2005}, the clustering coefficient of Type-I Apollonian network has been derived, and is equal to $0.83$. So, two types of Apollonian networks all display the cluster phenomena.

\begin{figure}
\centering
\includegraphics[height=7cm]{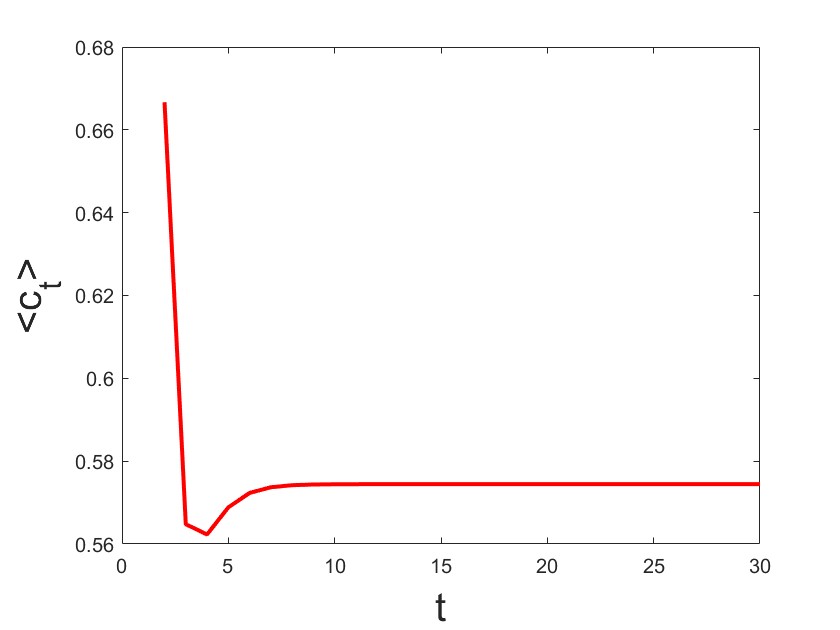}%\includegraphics[width=3in]{model2_ExpIncomeperTime_SymmetricVsAsymmetric.eps}
\caption{(Color online) The diagram of clustering coefficient $\langle c_{t}\rangle$ of network $\mathcal{A}_{t}$. The value for $\langle c_{t}\rangle$ first decreases at initial several time steps and after reaching minimal point it will gradually increase and finally tends to a stationary value, implying that model $\mathcal{A}_{t}$ has a higher clustering coefficient in the limit of large-$t$. }\label{fig:ExpIncomePerTime}
\end{figure}

\subsection{Diameter}

It is well known that small-world property on complex networks is often decided based on two structural parameters, namely clustering coefficient and diameter \cite{Newman-2003}. In the preceding subsection, we have studied clustering coefficient on network $\mathcal{A}_{t}$. In order to finally determine whether or not network $\mathcal{A}_{t}$ is small-world, it is necessary to consider the other structural parameter, i.e., diameter $\mathcal{D}_{t}$. In the language of graph theory \cite{Bondy-2008}, the diameter $\mathcal{D}$ of a graph $\mathcal{G}=(\mathcal{V},\mathcal{E})$ is defined as the total number of edges of a longest shortest path. To put this another way, diameter $\mathcal{D}$ is defined as $\mathcal{D}:= \text{max}\{d_{uv}:u,v\in\mathcal{V}\}$ in which $d_{uv}$ is the distance between vertices $u$ and $v$. It is clear to see that diameter is a global parameter for network $\mathcal{G}=(\mathcal{V},\mathcal{E})$. For instance, it is usually used to measure the delay of information on communication network as a whole. A smaller value corresponds to a better topological structure of network. In what follows, we evaluate diameter $\mathcal{D}_{t}$ on network $\mathcal{A}_{t}$.

\textbf{Proposition 4} The diameter $\mathcal{D}_{t}$ of network $\mathcal{A}_{t}$ is

\begin{equation}\label{eq:MF-2023-3-4-1}
D_{t}=t\propto \ln|\mathcal{V}_{t}|.
\end{equation}

Note that before beginning with our proof, Lemma 1 needs to be introduced, as follows.

\textbf{Lemma 1}  Given a pair of active vertices $u$ and $v$ in network $\mathcal{A}_{t}$ ($t\geq3$) that are not incident to each other, there is a path $P_{uv}(t)$ of length $d_{uv}(t)$ whose two end-edges are semi-active. Also, if distance $d_{uv}(t)$ is no less than $3$, there exists a cycle $C_{uv}(t)$ of length $2d_{uv}(t)$ that contains at least four semi-active end-edges.

\emph{Proof} We first consider the first portion. For a pair of active vertices $u$ and $v$ meeting the condition in Lemma 1, if distance $d_{uv}(t)$ is exactly equal to $2$, the pair of vertices must be connected to an identical vertex that is pre-existing. Clearly, the first portion in Lemma 1 holds in this case. As distance $d_{uv}(t)\geq3$, suppose that a given path $P_{uv}(t)$ of length $d_{uv}(t)$ does not comply with our requirement, we need to seek for another $P'_{uv}(t)$. Specifically, the desired path $P'_{uv}(t)$ can be obtained based on path $P_{uv}(t)$. For convenience, we denote by $uu_{1}u_{2}u$ that active triangle to which active vertex $u$ belongs and $x_{u}y_{u}z_{u}$ that prescribed triangle into which triangle $uu_{1}u_{2}u$ is added as described in Framework. In the meantime, active vertex $u$ is connected to vertices $x_{u}$ and $y_{u}$, active vertex $u_{1}$ is connected to vertices $z_{u}$ and $y_{u}$, and active vertex $u_{2}$ is connected to vertices $x_{u}$ and $z_{u}$. Then, without loss of generality, we assume that the path $P_{uv}(t)$ only contains an active end-edge, denoted by $e_{uu_{1}}$ for our purpose. Obviously, vertex $z_{u}$ belongs to path $P_{uv}(t)$. If no, replacing sub-path $uu_{1}y_{u}$ in path $P_{uv}(t)$ with edge $e_{uy_{u}}$ yields a new path that has length strictly smaller than $d_{uv}(t)$. This leads to a contradiction. Now, we replace sub-path $uu_{1}z_{u}$ in path $P_{uv}(t)$ with 2-path $ux_{u}z_{u}$, and then obtain a new path. It is clear to the eye that the resultant path has the same length as the given path $P_{uv}(t)$, and does not contain active end-edge at all. Therefore, this is a desired path $P'_{uv}(t)$. The other case, i.e., path $P_{uv}(t)$ with two active end-edges, is analyzed in a similar manner. This completes the first portion of Lemma 1.

Now, we consider the other portion. First, it is easy to verify that this portion holds on case of $t=3$ by listing out all possible cases. Next, we prove more general case by induction on parameter $t$. Suppose that this portion holds on case of $t-1$. We need to consider case of $t$. As above, we denote by $vv_{1}v_{2}v$ that active triangle to which active vertex $v$ belongs and $x_{v}y_{v}z_{v}$ that prescribed triangle into which triangle $vv_{1}v_{2}v$ is added as described in Framework. Meanwhile, active vertex $v$ is connected to vertices $x_{v}$ and $y_{v}$, active vertex $v_{1}$ is connected to vertices $z_{v}$ and $y_{v}$, and active vertex $v_{2}$ is connected to vertices $x_{v}$ and $z_{v}$. Then, without loss of generality, we assume that vertices $x_{u}$ and $x_{v}$ are active in the preceding model, i.e., network $\mathcal{A}_{t-1}$.  Based on assumption, there is a cycle $C_{x_{u}x_{v}}(t-1)$ of length $2d_{x_{u}x_{v}}(t-1)$. Clearly, edges $e_{x_{u}y_{u}}$, $e_{x_{u}z_{u}}$, $e_{x_{v}y_{v}}$ and $e_{x_{v}z_{v}}$ are semi-active in network $\mathcal{A}_{t-1}$, and thus belong to this cycle. For convenience, cycle $C_{x_{u}x_{v}}(t-1)$ is presented as $x_{u}y_{u}...y_{v}x_{v}z_{v}...z_{u}x_{u}$. Next, according to the first portion, we replace edge $e_{x_{u}y_{u}}$ in this cycle by 2-path $x_{u}uy_{u}$ and edge $e_{x_{v}z_{v}}$ in this cycle by 2-path $x_{v}vz_{v}$. The resulting cycle $x_{u}uy_{u}...y_{v}x_{v}vz_{v}...z_{u}x_{u}$ has length $2(d_{x_{u}x_{v}}(t-1)+1)$. As an immediate consequence, distance $d_{uv}(t)$ between vertices $u$ and $v$ is no more than $d_{x_{u}x_{v}}(t-1)+1$, namely, $d_{uv}(t)\leq d_{x_{u}x_{v}}(t-1)+1$. The left task to address is to show that distance $d_{uv}(t)$ is certainly equal to $d_{x_{u}x_{v}}(t-1)+1$. If no, assume that quantity $d_{uv}(t)$ is strictly smaller than $d_{x_{u}x_{v}}(t-1)+1$, i.e., $d_{uv}(t)\leq d_{x_{u}x_{v}}(t-1)$. According to this first portion, it is clear to see that the anticipated path $Q_{uv}(t)$ does not contain semi-active edges $e_{ux_{u}}$ and $e_{vx_{v}}$ simultaneously. Hence, two cases need to be discussed carefully.

\textbf{case 1} Path $Q_{uv}(t)$ contains two semi-active edges $e_{ux_{u}}$ and $e_{vz_{v}}$. If we remove semi-active edge $e_{ux_{u}}$ and replace edge $e_{vz_{v}}$ with edge $e_{z_{v}x_{v}}$, then the resulting path obviously connects vertices $x_{u}$ to $x_{v}$, and has length less than $d_{x_{u}x_{v}}(t-1)$. This leads to a contradiction.

\textbf{case 2} Path $Q_{uv}(t)$ contains two semi-active edges $e_{uy_{u}}$ and $e_{vx_{v}}$. By analogy with analysis in case 1, we also come to a contradiction.

According to the analysis above, we definitely state that distance $d_{uv}(t)$ is equal to $d_{x_{u}x_{v}}(t-1)+1$. This completes the second portion of Lemma 1. 

To sum up, Lemma 1 is completed. \qed

We are ready to verify Proposition 4. Now, let us start to provide a rigorous mathematical proof, as below.

\emph{Proof} First of all, it is easy to check that Eq.(\ref{eq:MF-2023-3-4-1}) holds on cases $t=1$ and $2$. From now on, let us rethink the evolution of network $\mathcal{A}_{t}$. From description shown in Framework, it is clear to the eye that for $t>2$, network $\mathcal{A}_{t}$ can in essence be constructed based on three networks $\mathcal{A}_{t-1}$. Hence, the relation $\mathcal{D}_{t}\geq\mathcal{D}_{t-1}$ certainly holds. In addition, the following expression is easy to consolidate \footnote{For two sets $A$ and $B$, symbol $A/B$ represents an induced set containing all elements $i$ that meets $i\in A$ and $i\notin B$.}.

\begin{equation}\label{eq:MF-2023-3-4-2}
D_{t}=\max\left\{d_{uv}(t):u,v\in \mathcal{V}_{t}/\mathcal{V}_{t-1}\right\}.
\end{equation}
Combining Lemma 1 and Eq.(\ref{eq:MF-2023-3-4-2}), we build up the following relation

\begin{equation}\label{eq:MF-2023-3-4-3}
D_{t}=\max\left\{d_{xy}(t-1):x,y\in \mathcal{V}_{t-1}/\mathcal{V}_{t-2}\right\}+1.
\end{equation}
Notice that the first term in the right hand side of Eq.(\ref{eq:MF-2023-3-4-3}) is by definition diameter of network $\mathcal{A}_{t-1}$. So, we gain

\begin{equation}\label{eq:MF-2023-3-4-4}
D_{t}=D_{t-1}+1,
\end{equation}
and finally derive $D_{t}=t$ in an iterative manner. This is complete. \qed

From Eqs.(\ref{eq:MF-2023-3-3-3}) and (\ref{eq:MF-2023-3-4-1}), we declare that the combination between smaller diameter and higher clustering coefficient indicates that Type-II Apollonian network $\mathcal{A}_{t}$ exhibits \emph{small-world property}, a characteristic that is prevalently observed in a lot of complex networks \cite{Watts-1998},\cite{Grabow-2015}, including Type-I Apollonian network \cite{Andrade-2005,Doye-2005}.

\subsection{Pearson correlation coefficient}

As the final structural parameter discussed in this section, Pearson correlation coefficient $r$ of network $\mathcal{G}=(\mathcal{V},\mathcal{E})$, firstly studied by Newman in \cite{Newman-2002}, is defined as
\begin{equation}\label{eqa:MF-2023-3-5-1}
r=\frac{|\mathcal{E}|^{-1}\sum\limits_{e_{ij}\in \mathcal{E}} k_{i}k_{j}-\left[|\mathcal{E}|^{-1}\sum\limits_{e_{ij}\in \mathcal{E}} \frac{1}{2}(k_{i}+k_{j})\right]^{2}}{|\mathcal{E}|^{-1}\sum\limits_{e_{ij}\in \mathcal{E}} \frac{1}{2}(k^{2}_{i}+k^{2}_{j})-\left[|\mathcal{E}|^{-1}\sum\limits_{e_{ij}\in \mathcal{E}} \frac{1}{2}(k_{i}+k_{j})\right]^{2}},
\end{equation}
in which $k_{i}$ is degree of vertex $i$ and $e_{ij}$ denotes an edge connecting vertex $i$ to $j$. Newman has proven both empirically and analytically that most social networks belong to the assortative mixing family ($r>0$) and, however, almost all biological and technological networks fall into the scope of disassortative mixing ($r<0$) \cite{Newman-2002}.

By definition, one can clearly understand that parameter $r$ measures tendency of connections taking place between vertices of network $\mathcal{G}=(\mathcal{V},\mathcal{E})$. Case of $r>0$ means that there is a clear preference for vertex to link to other vertex like itself with regard of vertex degree. And, case of $r<0$ indicates the opposite consequence. It is worth noticing that there is a critical case $r=0$. In the meantime, network in question might be a regular graph. Bear this in mind, let us start to calculate the exact solution of such a parameter on network $\mathcal{A}_{t}$.

\textbf{Proposition 5} The Pearson correlation coefficient $r_{t}$ of network $\mathcal{A}_{t}$ is

\begin{equation}\label{eqa:MF-2023-3-5-2}
r_{t}=\frac{\frac{2\Gamma_{t}(1)}{3^{t+1}-3}-\left[\frac{\Gamma_{t}(2)}{3^{t+1}-3}\right]^{2}}
{\frac{\Gamma_{t}(1)}{3^{t+1}-3}-\left[\frac{\Gamma_{t}(2)}{3^{t+1}-3}\right]^{2}}
\end{equation}
here

\begin{subequations}
\label{eq:whole}
\begin{eqnarray}
\begin{aligned}\Gamma_{t}(1)=192\times3^{t-2}+\sum_{i=0}^{t-3}3^{i}&\left([3(t-i)+9]2^{2(t-i)-1}\right.\\
&\left.+[3(t-i)+6]2^{t-i}-12 \right),
\end{aligned}\label{subeq:MF-1-1}
\end{eqnarray}
\begin{equation}
\begin{aligned}\Gamma_{t}(2)=96\times3^{t-2}+\sum_{i=0}^{t-3}3^{i}&\left(3\times2^{2(t-i)-1}\right.\\
&\left.+15\times2^{t-i-1}+12 \right),
\end{aligned}\label{subeq:MF-1-2}
\end{equation}
\begin{equation}
\begin{aligned}\Gamma_{t}(3)=384\times3^{t-2}+\sum_{i=0}^{t-3}3^{i}&\left(9\times2^{3(t-i)-2}+9\times2^{2(t-i)-1}\right.\\
&\left.+3\times2^{(t-i)+3}+24\right).
\end{aligned}\label{subeq:MF-1-3}
\end{equation}
\end{subequations}

\emph{Proof} Before we dive into more details, some notations need to be introduced. Among them, we denote by $h_{t}$ the greatest degree vertex in network $\mathcal{A}_{t}$. From topological structure of network $\mathcal{A}_{t}$, one can see that there are three vertices $h^{i}_{t}$ ($i=1,2,3$). At the same time, we make use of notation $\mathcal{M}_{h_{t}}$ to indicate the neighboring set of vertex $h_{t}$. Obviously, an arbitrary set $\mathcal{M}_{h_{t}}$ contains the other two vertices $h_{t}$. If we remove these two $h_{t}$s, the resulting set is denoted by $\mathcal{M}^{\ddagger}_{h_{t}}$. In addition, we simplify the presentation in Eq.(\ref{eqa:MF-2023-3-5-1}) by using the following notations

$$\Gamma_{t}(1)=\sum_{e_{uv}\in \mathcal{E}_{t}}k_{u}k_{v}, \quad \Gamma_{t}(2)=\sum_{e_{uv}\in \mathcal{E}_{t}}(k_{u}+k_{v}),\quad \text{and} $$ $$\Gamma_{t}(3)=\sum_{e_{uv}\in \mathcal{E}_{t}}(k^{2}_{u}+k^{2}_{v})$$
when studying network $\mathcal{A}_{t}$.

We are now ready to derive the precise solution to Pearson correlation coefficient $r_{t}$ of network $\mathcal{A}_{t}$. According to the topological structure of network $\mathcal{A}_{t}$, quantity $\Gamma_{t}(1)$ is given by

\begin{equation}\label{eqa:MF-2023-3-5-3}
\begin{aligned}\Gamma_{t}(1)&=3\Gamma_{t-1}(1)+6\sum_{u\in \mathcal{N}^{\ddagger}_{h_{t-1}}}k_{h_{t-1}}k_{u}+3\times3k^{2}_{h_{t-1}}\\
&+6(k^{2}_{h_{t-1}}+4k_{h_{t-1}})+3(k_{h_{t-1}}+2)^{2}+3\sum_{u\in \mathcal{N}^{\ddagger}_{h_{t-1}}}2k_{u}\\
&=3\Gamma_{t-1}(1)+(6\times2^{t-1}+6)(\Phi_{t-1}-2\times2^{t-1})\\
&+9\times2^{2(t-1)}+6(2^{2(t-1)}+2^{t+1})+3(2^{t-1}+2)^{2}\\
&=3\Gamma_{t-1}(1)+(3t+9)2^{2t-1}+(3t+6)2^{t}-12.
\end{aligned}
\end{equation}
Note that we have used $\Phi_{t}=\sum_{u\in \mathcal{M}_{h_{t}}}k_{u}=(t+3)2^{t}-4$. Based on Eq.(\ref{eqa:MF-2023-3-5-3}), one can obtain Eq.(\ref{subeq:MF-1-1}) by using an iterative calculation. Similarly, we obtain a group of equations as follows

\begin{equation}\label{eqa:MF-2023-3-5-4}
\begin{aligned}\Gamma_{t}(2)&=3\Gamma_{t-1}(2)+6\sum_{u\in \mathcal{N}^{\ddagger}_{h_{t-1}}}k_{h_{t-1}}+3\times3k_{h_{t-1}}\\
&+6(k_{h_{t-1}}+2)+3\times2(k_{h_{t-1}}+2)+3\sum_{u\in \mathcal{N}^{\ddagger}_{h_{t-1}}}2\\
&=3\Gamma_{t-1}(2)+6\times2^{t-1}(2^{t-1}-2)+9\times2^{t-1}\\
&+6\times2^{t-1}+12+3\times2^{t}+12+6(2^{t-1}-2)\\
&=3\Gamma_{t-1}(2)+3\times2^{2t-1}+15\times2^{t-1}+12,
\end{aligned}
\end{equation}
and
\begin{equation}\label{eqa:MF-2023-3-5-5}
\begin{aligned}\Gamma_{t}(3)&=3\Gamma_{t-1}(3)+6\sum_{u\in \mathcal{N}^{\ddagger}_{h_{t-1}}}3k^{2}_{h_{t-1}}+3\times6k^{2}_{h_{t-1}}\\
&+6(3k^{2}_{h_{t-1}}+4k_{h_{t-1}}+4)+3\times2(k_{h_{t-1}}+2)^{2}\\
&+3\sum_{u\in \mathcal{N}^{\ddagger}_{h_{t-1}}}(4k_{h_{t-1}}+4)\\
&=3\Gamma_{t-1}(3)+18\times2^{2(t-1)}(2^{t-1}-2)+18\times2^{2(t-1)}\\
&+6(3\times2^{2(t-1)}+2^{t+1}+4)+6(2^{t-1}+2)^{2}\\
&\quad+3(2^{t+1}+4)(2^{t-1}-2)\\
&=3\Gamma_{t-1}(3)+9\times2^{3t-2}+9\times2^{2t-1}+3\times2^{t+3}+24.
\end{aligned}
\end{equation}
It should be mentioned that the initial conditions $\Gamma_{2}(1)=192$, $\Gamma_{2}(2)=96$ and $\Gamma_{2}(3)=384$ need to be used when deriving the exact solutions of quantities $\Gamma_{t}(1)$, $\Gamma_{t}(2)$ and $\Gamma_{t}(3)$. Armed with these results, we confirm that Eq.(\ref{eqa:MF-2023-3-5-2}) holds after some simple arithmics. This is complete. \qed

To show the scaling of Pearson correlation coefficient $r_{t}$ in the large graph size limit, we feed network $\mathcal{A}_{t}$ into computer and gain an illustrative outline in Fig.6. Obviously, the value for $r_{t}$ is constantly negative and approaches $0$ in the limit of large-$t$, which implies that network $\mathcal{A}_{t}$ possesses \emph{disassortative structure}. This is similar to observation on Type-I Apollonian network. In another word, these two types of Apollonian networks are disassortative.

\begin{figure}
\centering
\includegraphics[height=7cm]{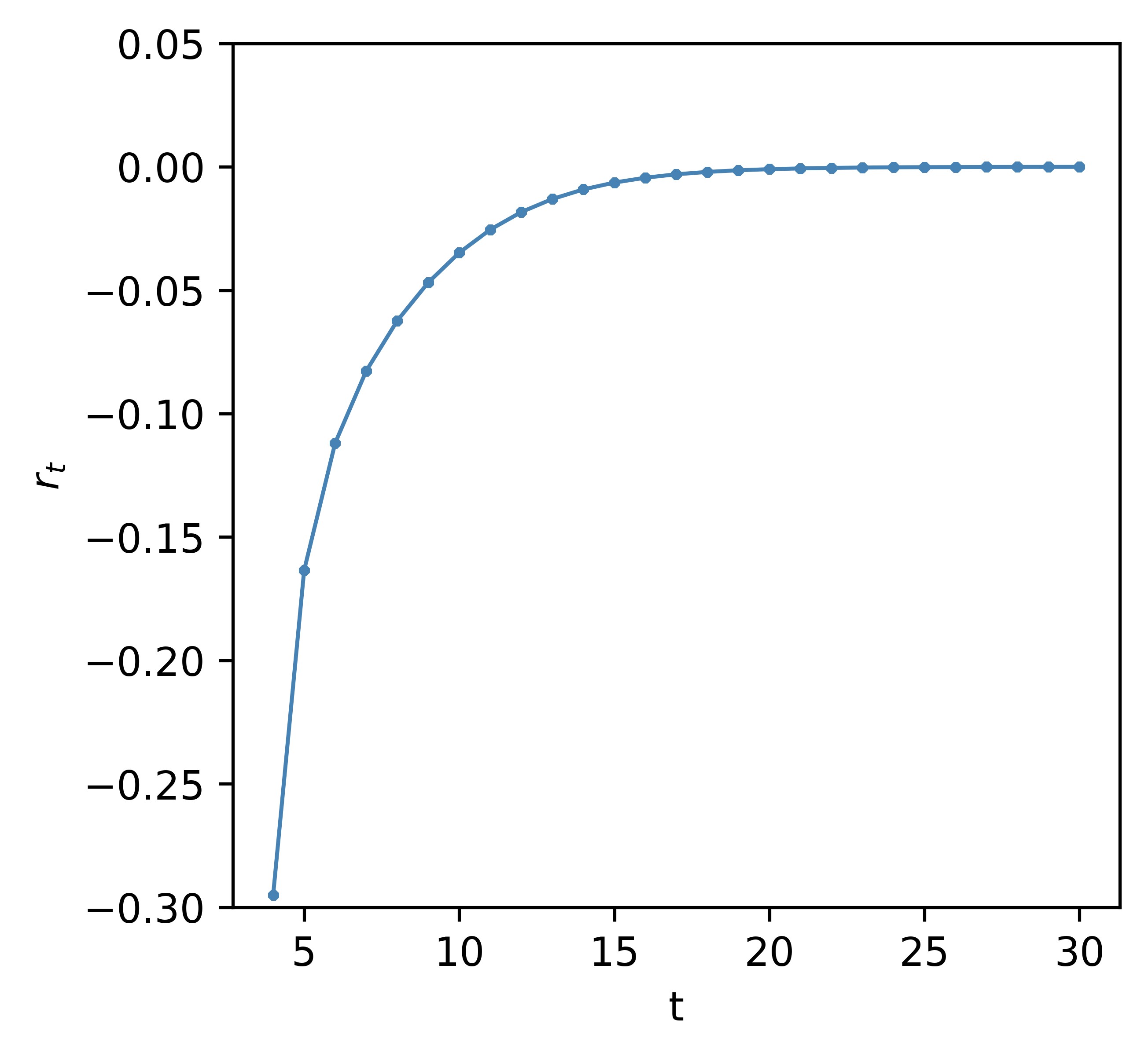}%\includegraphics[width=3in]{model2_ExpIncomeperTime_SymmetricVsAsymmetric.eps}
\caption{(Color online) The diagram of Pearson correlation coefficient $r_{t}$ of network $\mathcal{A}_{t}$. Clearly, the negative value for $r_{t}$ means that the proposed model $\mathcal{A}_{t}$ is disassortative.}\label{fig:ExpIncomePerTime}
\end{figure}

Until now, we study some basic structural parameters on network $\mathcal{A}_{t}$ in more detail. Due to these consequences above, we firmly demonstrate that the proposed model $\mathcal{A}_{t}$ is sparse, follows power-law degree distribution and has small-world property. At the same time, model $\mathcal{A}_{t}$ possesses disassortative structure due to Pearson correlation coefficient $r_{t}$ smaller than zero. Besides that, network $\mathcal{A}_{t}$ must have some other structural features of considerable interest. For the sake of the outline of this paper, we do not plan to enter into more details. They are left as our next move. Interested reader may also probe them. In the next section, we will discuss the problem of how to enumerate spanning trees on model $\mathcal{A}_{t}$, and derive the solution to the spanning tree entropy.

\section{Spanning trees}

Given a graph $\mathcal{G}=(\mathcal{V},\mathcal{E})$, its spanning subgraph $\mathcal{G}'=(\mathcal{V},\mathcal{E}')$ is a sub-graph with the same vertex set as $\mathcal{G}=(\mathcal{V},\mathcal{E})$ and a number of edges $\mathcal{E}'$ such that $|\mathcal{E}'|\leq|\mathcal{E}|$. A spanning tree $\mathcal{T}=(\mathcal{V},\mathcal{E}')$ of a connected graph $\mathcal{G}=(\mathcal{V},\mathcal{E})$ is a spanning subgraph which is a tree having $|\mathcal{E}'|=|\mathcal{V}|-1$. The spanning trees number on graph $\mathcal{G}=(\mathcal{V},\mathcal{E})$ is an important structural parameter in the filed of graph theory and theoretical computer science \cite{Ma-TCS-2018}-\cite{Li-2021}. It always plays a key role not just on understanding some structural features, but also on determining some relevant dynamical properties, including instance reliability \cite{G-J-S-2003}, synchronization capability \cite{N-T-A-E-2006,Afshari-2020}, random walks \cite{P-M-2000,Alev-2020}, to name just a few.

It is widely known that the problem of calculating the total number of spanning trees of any finite graph had been theoretically addressed by the well-known Kirchhoff's matrix-tree theorem. Specifically speaking, the exact solution of spanning tree number of a graph is equivalent to the product of all nonzero eigenvalues of the Laplacian matrix of graph in question. When we consider an enormous network model with hundreds and thousands of vertices and edges, such a theoretical accomplishment might not be easy to implement only because of the huge overhead of computing resources. Hence, lots of works related to determination of the number of spanning trees of special network models need to be done in the future. For the proposed network $\mathcal{A}_{t}$, we will provide a rigorous mathematical approach to count spanning trees. To make further progress, we design an effective algorithm for calculating the exact solution of spanning trees number of network $\mathcal{A}_{t}$.

\begin{figure}
\centering
\includegraphics[height=5cm]{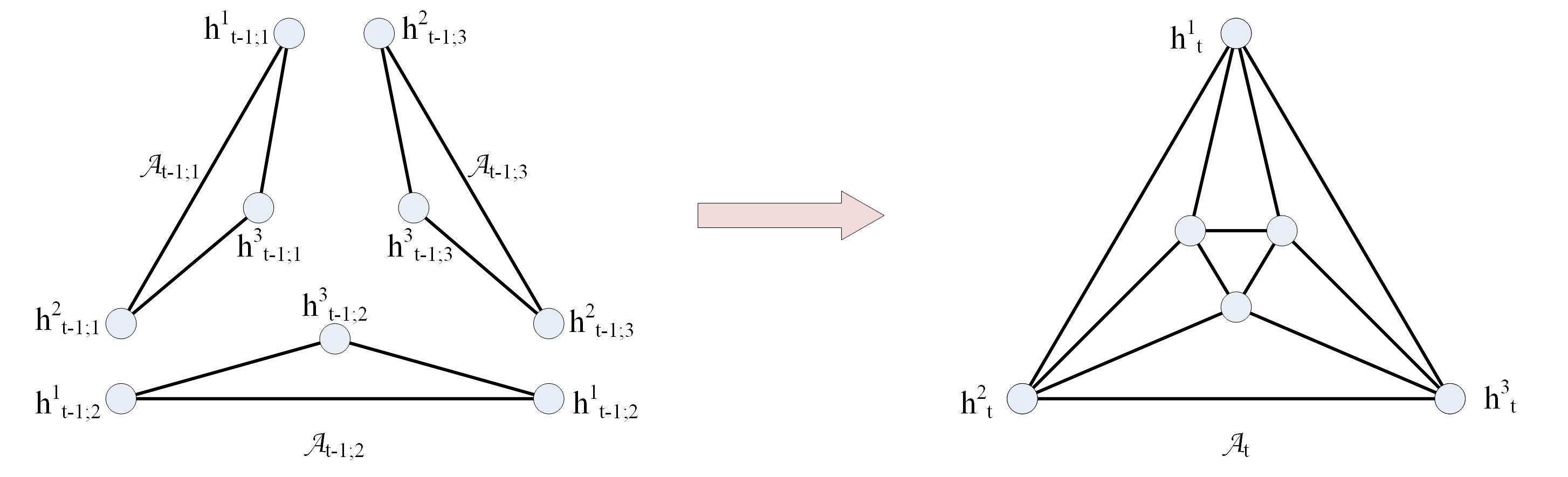}%\includegraphics[width=3in]{model2_ExpIncomeperTime_SymmetricVsAsymmetric.eps}
\caption{(Color online) The diagram of reconstruction of network $\mathcal{A}_{t}$ based on three networks $\mathcal{A}_{t-1,i}$ where $i=1,2,3$. }\label{fig:ExpIncomePerTime}
\end{figure}

To this end, let us rethink how network $\mathcal{A}_{t}$ evolves over time. Clearly, there exist three greatest degree vertices in network $\mathcal{A}_{t}$. We denote by $h^{i}_{t}$ these greatest degree vertices in network $\mathcal{A}_{t}$ as used in the section above. See Fig.7 for an illustrative example. From description in Framework, it is also clear to the eye that as $t\geq2$, network $\mathcal{A}_{t}$ can in essence be constructed based on three networks $\mathcal{A}_{t-1}$. More specifically, given three networks $\mathcal{A}_{t-1;j}$ where $j=1,2,3$, we can reconstruct network $\mathcal{A}_{t}$ by not only identifying vertices $h^{1}_{t-1;1}$ and $h^{2}_{t-1;3}$ into vertex $h^{1}_{t}$, vertices $h^{2}_{t-1;1}$ and $h^{1}_{t-1;2}$ into vertex $h^{2}_{t}$ as well as vertices $h^{2}_{t-1;2}$ and $h^{1}_{t-1;3}$ into vertex $h^{3}_{t}$, but also adding three new edges between vertices $h^{3}_{t-1;1}$, $h^{3}_{t-1;2}$ and $h^{3}_{t-1;3}$ so as to generate a triangle. In fact, these newly added edges are active as defined in remark. This kind of construction presented here is called \emph{self-construct growth} of network $\mathcal{A}_{t}$ for our purpose, which is used later.

On the basis of both the construction of network $\mathcal{A}_{t}$ and the definition of spanning tree, we introduce some notations. Given network $\mathcal{A}_{t}$, the $h^{i}_{t}$-$h^{j}_{t}$ $2$-spanning-forest $\mathcal{F}_{h^{i}_{t}\wedge h^{j}_{t}}(t)$ is a spanning subgraph, and consists of two trees in which one contains vertex $h^{i}_{t}$ and the other contains vertex $h^{j}_{t}$. Given vertex pair $h^{i}_{t}$ and $h^{j}_{t}$, the total number of $2$-spanning-forests of such type is indicated by notation $b_{t;i\wedge j}$. The $h^{1}_{t}$-$h^{2}_{t}$-$h^{3}_{t}$ $3$-spanning-forest $\mathcal{F}_{h^{1}_{t}\wedge h^{2}_{t}\wedge h^{3}_{t}}(t)$ is also a spanning subgraph, but consists of three trees each of which only contains one vertex $h^{i}_{t}$. The total number of $3$-spanning-forests of such kind is denoted by notation $c_{t}$. From the concrete description in Framework, it is obvious to understand that three quantities $b_{t;i\wedge j}$ are equivalent to each other. For brevity, we identify them into quantity $b_{t}$, i.e., $b_{t}:=b_{t;i\wedge j}$ where $i,j=1,2,3$ and $i\neq j$, which is used in the rest of this paper. At last, we use $s_{t}$ to represent the number of spanning trees of network $\mathcal{A}_{t}$.

We now establish an effective algorithm, called Algorithm 1 for convenience, in order to exactly determine the solution of spanning trees number of network $\mathcal{A}_{t}$. First of all, we give a lemma for the purpose of ensuring the rationality and validity of Algorithm 1.

\textbf{Lemma 2} As $t\geq2$, quantities $s_{t}$, $b_{t}$ and $c_{t}$ in network $\mathcal{A}_{t}$ obey

\begin{equation}\label{eq:MF-2023-4-2-1}
\left\{\begin{aligned}&s_{t+1}=6s_{t}^{2}b_{t}+6s_{t}^{2}c_{t}+42s_{t}b_{t}^{2}+36s_{t}b_{t}c_{t}+42b_{t}^{3}\\
&b_{t+1}=4s_{t}b_{t}^{2}+3s_{t}c_{t}^{2}+8s_{t}b_{t}c_{t}+12b_{t}^{2}c_{t}+14b_{t}^{3}\\
&c_{t+1}=8b_{t}^{3}+24b_{t}^{2}c_{t}+18b_{t}c_{t}^{2}
\end{aligned}\right.
\end{equation}
where the initial conditions $s_{1}$, $b_{1}$ and $c_{1}$ are given by $s_{1}=3$, $b_{1}=1$, $c_{1}=1$.

\emph{Proof} We only prove the correctness of expression in the first line of Eq.(\ref{eq:MF-2023-4-2-1}), the other two can be verified in a similar manner. Below is a detailed presentation, which is based on self-construct growth of network $\mathcal{A}_{t}$ (also see Figs.9 and 10 in Supplementary Materials for more information).

\textbf{case 1} According to definition of spanning tree and self-construct growth of network $\mathcal{A}_{t}$, it is impossible that an arbitrary spanning tree of network $\mathcal{A}_{t+1}$ is obtained by three small trees each of which is a spanning tree of network $\mathcal{A}_{t}$.

\textbf{case 2} It is possible that a spanning tree of network $\mathcal{A}_{t+1}$ is obtained by three small components\footnote{Here, each component is from a designated network $\mathcal{A}_{t}$. As shown in Fig.7, network $\mathcal{A}_{t+1}$ is constructed by three networks $\mathcal{A}_{t}$. This kind of statement is used in the rest of the proof.}. Among them, two are spanning trees of network $\mathcal{A}_{t}$ and the left is a $2$-spanning-forest. In this case, we obtain $6s_{t}^{2}b_{t}$ spanning trees in total. Furthermore, the left might also be a $3$-spanning-forest. As a consequence, we have $6s_{t}^{2}c_{t}$ spanning trees again. See the first line of Fig.9 in Supplementary Materials for an illustrative explanation.

\textbf{case 3} Similarly, it is also possible that a spanning tree of network $\mathcal{A}_{t+1}$ is obtained by three small components. Among them, one is a spanning tree of network $\mathcal{A}_{t}$ and the left two are $2$-spanning-forests. We enumerate all possible combinations in this case, and obtain $42s_{t}b^{2}_{t}$ spanning trees.

\textbf{case 4} If three small components are distinct with each other, then it is also possible that we construct a spanning tree of network $\mathcal{A}_{t+1}$ based on these components. The total number of spanning trees of this kind is calculated to equal $36s_{t}b_{t}c_{t}$.

\textbf{case 5} Suppose that three given components are $2$-spanning-forests of network $\mathcal{A}_{t}$, it turns out to be possible that a spanning tree of network $\mathcal{A}_{t+1}$ can be obtained by means of these $2$-spanning-forests. Notice that in this case, we have $42b_{t}^{3}$ all possible combinations, and thus gain $42b_{t}^{3}$ spanning trees.

Other cases not listed here can not leads to a spanning tree of network $\mathcal{A}_{t+1}$. Hence, according to cases1-5, the total number of spanning trees of network $\mathcal{A}_{t+1}$ is given by

$$s_{t+1}=6s_{t}^{2}b_{t}+6s_{t}^{2}c_{t}+42s_{t}b_{t}^{2}+36s_{t}b_{t}c_{t}+42b_{t}^{3}.$$

Although we omit the detailed proof of quantities $b_{t}$ and $c_{t}$, Figs.11 and 12 in Supplementary Materials show graphic presentation that is helpful to understand more details.

By definition, it is easy to derive the exact solutions of quantities $s_{1}$, $b_{1}$ and $c_{1}$. This completes Lemma 2. \qed

With the help of Lemma 2, we propose Algorithm 1 as below. It should be mentioned that the exact solution of spanning tree number $s_{t}$ of network $\mathcal{A}_{t}$ is obtained based on Eq.(\ref{eq:MF-2023-4-2-1}). Interested reader may make an attempt.  

\begin{algorithm}
	\caption{Counting the total number of spanning trees on Type-II Apollonian network $\mathcal{A}_{t}$.}  
	\label{alg:Framwork} 
	\begin{algorithmic}[1]
		\Require  
		network $\mathcal{A}_{t}$;
		initial conditions $s_{1}=3$, $b_{1}=1$ and $c_{1}=1$;
	    parameter $t$
		\Ensure  
		$s_{t}$
		\For{$i = \{1,2,...,t-1\}$} 
			\State
				$s_{i+1}\leftarrow 6s_{i}^{2}b_{i}+6s_{i}^{2}c_{i}+42s_{i}b_{i}^{2}+36s_{i}b_{i}c_{i}+42b_{i}^{3}$
			\State
				$b_{i+1} \leftarrow 4s_{i}b_{i}^{2}+3s_{i}c_{i}^{2}+8s_{i}b_{i}c_{i}+12b_{i}^{2}c_{i}+14b_{i}^{3}$
			\State
				$c_{i+1} \leftarrow 8b_{i}^{3}+24b_{i}^{2}c_{i}+18b_{i}c_{i}^{2}$
		\EndFor
		\State
		\Return $s_t$
	\end{algorithmic}  
\end{algorithm}

In \cite{Lyons-2005}, the spanning tree entropy, denoted by $\xi$, of a graph $\mathcal{G}=(\mathcal{V},\mathcal{E})$ is defined as 

$$\xi=\frac{\ln s}{|\mathcal{V}|},$$
where $s$ represents the total number of spanning trees.

Based on this, we have the following proposition. 

\textbf{Proposition 6} The spanning tree entropy of network $\mathcal{A}_{t}$ is given by
\begin{equation}\label{eq:MF-2023-4-1}
\xi_{\mathcal{A}}=\lim_{t\rightarrow \infty}\xi_{t}=\lim_{t\rightarrow \infty}\frac{\ln s_{t}}{|\mathcal{V}_{t}|} =1.44.
\end{equation}
This is easily verified by virtue of Eq.(\ref{eq:MF-2023-2-2}) and Algorithm 1, we hence omit proof. It is worth noticing that Fig.8 shows the tendency of quantity $\xi_{t}$ in the large-$t$ limit.

In \cite{Zhang-2014,Zhang-2013}, the spanning tree entropy of Type-I Apollonian network is calculated to asymptotically equal 1.35. This suggests that Type-II Apollonian network $\mathcal{A}_{t}$, as it has higher the spanning tree entropy, is more reliable to a random
removal of edges than Type-I Apollonian network.

\begin{figure}
\centering
\includegraphics[height=7cm]{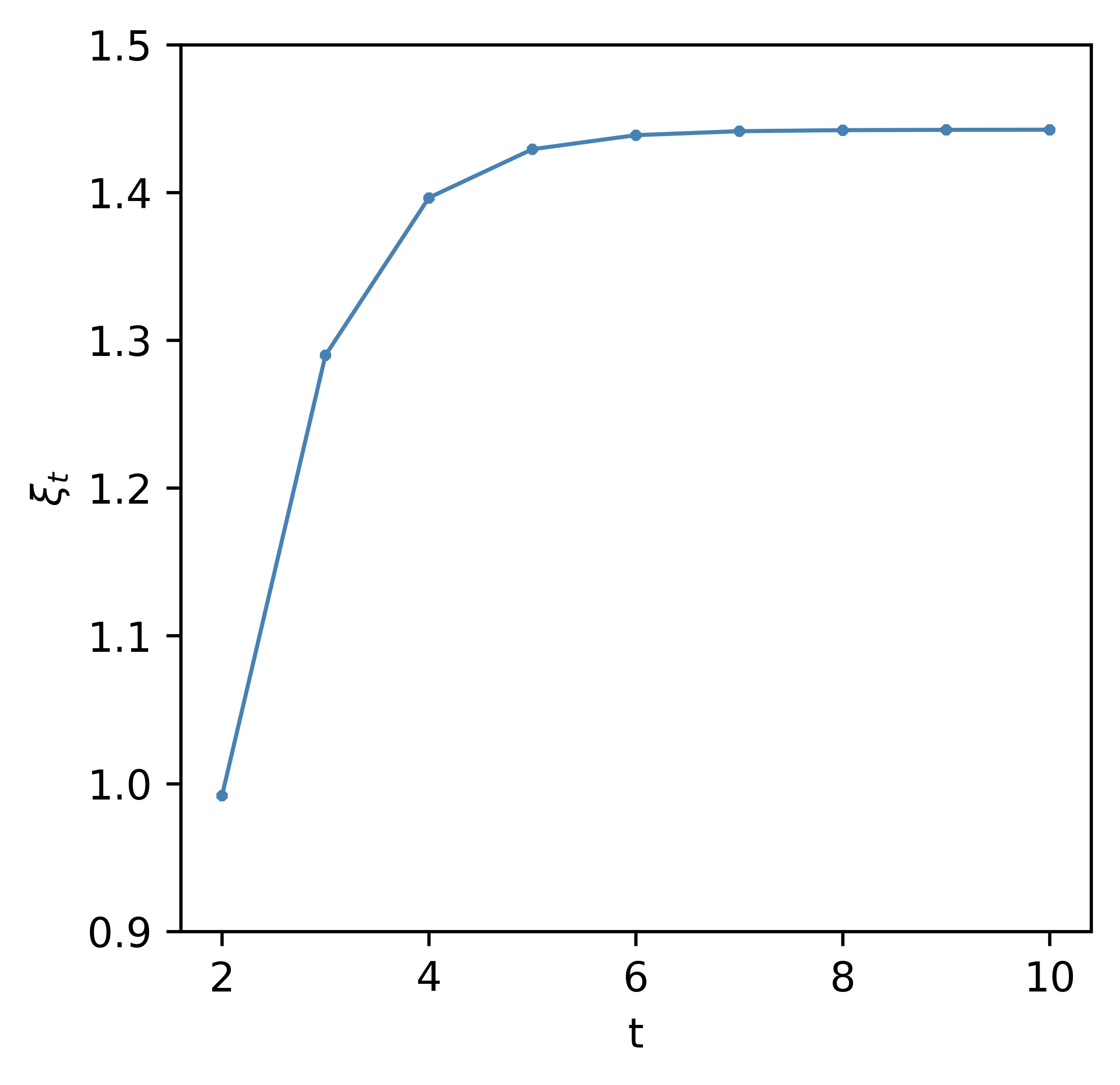}%\includegraphics[width=3in]{model2_ExpIncomeperTime_SymmetricVsAsymmetric.eps}
\caption{(Color online) The spanning tree entropy of network $\mathcal{A}_{t}$. }\label{fig:ExpIncomePerTime}
\end{figure}

\section{Trapping problem}

In this section, we introduce trapping problem on network $\mathcal{G}=(\mathcal{V},\mathcal{E})$. As mentioned above, it is a specific case of unbiased discrete-time random walk. So, we first need to present some details about random walks on network $\mathcal{G}=(\mathcal{V},\mathcal{E})$ \cite{Noh-2004}. Specifically speaking, a walker who performs random walks moves from its current location $u$ with a uniform probability $1/k_{u}$ to each candidate $v$ in its neighboring vertex set. In the jargon of mathematics, this kind of stochastic process can be precisely described by the transition matrix $\mathbf{P}_{\mathcal{G}}=\mathbf{D}_{\mathcal{G}}^{-1}\mathbf{A}_{\mathcal{G}}$ whose entry $p_{uv}=a_{uv}/k_{u}$ indicates the probability of jumping from $u$ to $v$ in one step. Here, $\mathbf{A}_{\mathcal{G}}$ and $\mathbf{D}_{\mathcal{G}}$ represent, respectively, adjacency matrix and diagonal degree matrix. That is to say, $\mathbf{A}_{\mathcal{G}}=\left(a_{uv}\right)$ where $a_{uv}$ is equal to $1$ if vertex $u$ is adjacent to vertex $v$, and $0$ otherwise. $\mathbf{D}_{\mathcal{G}}$ is given by $\mathbf{D}_{\mathcal{G}}=\text{diag}(k_{1},k_{2},...,k_{|\mathcal{V}|})$. $\mathbf{D}_{\mathcal{G}}^{-1}$ denotes the inverse of matrix $\mathbf{D}_{\mathcal{G}}$. Analogously, we obtain the normalized Laplacian matrix $\mathbf{L}_{\mathcal{G}}=\mathbf{I}_{\mathcal{G}}-\mathbf{P}_{\mathcal{G}}$. 

In the trapping problem on network $\mathcal{G}=(\mathcal{V},\mathcal{E})$, a trap is positioned at a given location (in fact, a vertex $u$) on network that absorbs all particles to visit it by performing random walks \cite{Montroll-1969}. One of most fundamental parameters pertaining to trapping problem is trapping time $\mathcal{T}_{v\rightarrow u}$ that is defined as the expected time taken by a walker starting out from source vertex $v$ to first hit destination vertex $u$ (namely, trap). For the whole network $\mathcal{G}=(\mathcal{V},\mathcal{E})$, average trapping time $\langle\mathcal{T}_{\mathcal{G}}\rangle$ is defined as

\begin{equation}\label{eqa:MF-5-0-1}
\langle\mathcal{T}_{\mathcal{G}}\rangle=\frac{\sum_{v(\neq u)\in\mathcal{V}_{\mathcal{G}}}\mathcal{T}_{v\rightarrow u}}{|\mathcal{V}_{\mathcal{G}}|-1}.
\end{equation}
Hence, it suffices to derive the solution of trapping time $\mathcal{T}_{v\rightarrow u}$ in order to obtain quantity $\langle\mathcal{T}_{\mathcal{G}}\rangle$. For ease of presentation, we label vertex $u$, which is assigned the trap, with number $1$. Similarly, other vertices in network $\mathcal{G}=(\mathcal{V},\mathcal{E})$ are distinguished using a unique number $i(\in[2,|\mathcal{V}|])$.  

Due to the nature of Markov Chain \cite{Haggstrom-2002}, it is clear to see that trapping time $\mathcal{T}_{v\rightarrow u}$ follows the coming equation 

\begin{equation}\label{eqa:MF-5-0-2}
\mathcal{T}_{v\rightarrow u}=\sum_{w(\neq u)\in\mathcal{V}_{\mathcal{G}}}\frac{a_{vw}}{k_{v}}\mathcal{T}_{w\rightarrow u}+1.
\end{equation}  
Now, all quantities $\mathcal{T}_{v\rightarrow u}$ are collected to yield a vector $\mathbf{T}_{\mathcal{G}}=(\mathcal{T}_{2\rightarrow u},...,\mathcal{T}_{|\mathcal{V}|\rightarrow u})$. From Eq.(\ref{eqa:MF-5-0-2}), vector $\mathbf{T}_{\mathcal{G}}$ is expressed as follows 

\begin{equation}\label{eqa:MF-5-0-3}
\mathbf{T}_{\mathcal{G}}=\overline{\mathbf{L}_{\mathcal{G}}}^{-1}\mathbf{e}^{\top},
\end{equation} 
where $\overline{\mathbf{L}_{\mathcal{G}}}$ is an induced submatrix by deleting the row and column associated with vertex $u$ of matrix $\mathbf{L}_{\mathcal{G}}$, and $\mathbf{e}$ is the $|\mathcal{V}|-1$-dimensional all-one vector, i.e., $\mathbf{e}=(1,1,...,1)$. Symbol $\top$ indicates transpose of matrix. Based on the statement above, Eq.(\ref{eqa:MF-5-0-1}) is rewritten as 

\begin{equation}\label{eqa:MF-5-0-4}
\langle\mathcal{T}_{\mathcal{G}}\rangle=\frac{\sum_{i=2}^{|\mathcal{V}|}\sum_{j=2}^{|\mathcal{V}|}\overline{L_{ij}}}{|\mathcal{V}_{\mathcal{G}}|-1}, 
\end{equation}
in which $\overline{L_{ij}}$ is entry of matrix $\overline{\mathbf{L}_{\mathcal{G}}}^{-1}$, and $\sum_{j=2}^{|\mathcal{V}|}\overline{L_{ij}}$ is by definition quantity $\mathcal{T}_{i\rightarrow u}$. This suggests that the problem of deriving average trapping time $\langle\mathcal{T}_{\mathcal{G}}\rangle$ is reduced to determining the summation of all entries of matrix $\overline{\mathbf{L}_{\mathcal{G}}}^{-1}$. In essence, this kind of calculation is a general manner in which quantity $\langle\mathcal{T}_{\mathcal{G}}\rangle$ for an arbitrary graph can be obtained. 

From now on, let us begin to study trapping problem on Type-II Apollonian network $\mathcal{A}_{t}$. As mentioned above, we also select the hub as location on which trap is allocated. Note that two distinct settings will be discussed in more detail. In the first setting, walker on each vertex $v$ performs random walk independently. This is called vertex-based strategy for our purpose. However, we suppose that in the other setting, walkers on some vertices will perform random walks by virtue of some consensus. More details are contained in the coming subsections.

\subsection{Vertex-based strategy}

We are now ready to consider trapping problem on Type-II Apollonian network $\mathcal{A}_{t}$ in the first setting. It should be mentioned that the hubs in network $\mathcal{A}_{t}$ are designed traps. From the concrete construction of network $\mathcal{A}_{t}$ in Section 2, it follows that there are three hubs in network $\mathcal{A}_{t}$. For convenience, we employ notation used in the above section, and denote by $h_{t}$ three hubs. In addition, we need to introduce some notations used below. Each vertex in network $\mathcal{A}_{t}$ is assigned a unique label. Specifically, we label each vertex added into network $\mathcal{A}_{t}$ at time step $i$ utilizing a unique integer in range $[|\mathcal{V}_{i-1}|+1,|\mathcal{V}_{i}|]$ where $i$ is equal to $1,2,...,t$. An illustrative example is shown in Fig.13 in Supplementary Materials.    

\textbf{Proposition 7} In the large graph size limit, i.e., $t\rightarrow\infty$, the average trapping time $\langle\mathcal{T}_{t}\rangle$ on network $\mathcal{A}_{t}$ follows
\begin{equation}\label{eq:MF-5-1-0}
\langle\mathcal{T}_{t}\rangle<O\left(|\mathcal{V}_{t}|^{2-\frac{\ln5}{\ln3}}\right).
\end{equation}

\emph{Proof} Before beginning with the detailed calculations, let us consider two simple examples. In network $\mathcal{A}_{2}$, on the basis of Eq.(\ref{eqa:MF-5-0-4}), it is easy to obtain the following equation 

\begin{equation}\label{eqa:MF-5-1-1}
\mathcal{T}_{i\rightarrow h_{2}}(2)=2,\quad i=4,5,6. 
\end{equation}
Note that $\mathcal{T}_{i\rightarrow h_{2}}(2)$ is trapping time for a walker starting out from vertex $i$ in network $\mathcal{A}_{2}$. It is without difficulty that in network $\mathcal{A}_{3}$, we take advantage of a similar technique to obtain 

\begin{equation}\label{eqa:MF-5-1-2}
\left\{\begin{aligned}&\mathcal{T}_{i\rightarrow h_{3}}(3)=\frac{25}{8},\quad i=4,5,6;\\
&\mathcal{T}_{i\rightarrow h_{3}}(3)=\frac{13}{4},\quad i=7,...,12;\\
&\mathcal{T}_{i\rightarrow h_{3}}(3)=\frac{21}{8},\quad i=13,14,15.
\end{aligned}\right. 
\end{equation}   

As mentioned above, the technique based on Eq.(\ref{eqa:MF-5-0-4}) is general, but becomes more complicated or prohibitive to implement when the number of vertices of network $\mathcal{A}_{t}$ increases exponentially with time $t$. On the other hand, it is of great importance to evaluate the scale of quantity one is interested in when discussing the finite-size effect in the scaling behavior. Therefore, we will make use of another method to determine the scale of average trapping time for network $\mathcal{A}_{t}$ in the limit of large graph size. This is proceeded mainly based on the following findings.

Due to Eq.(\ref{eqa:MF-5-1-2}), we come to the coming relationship

\begin{equation}\label{eqa:MF-5-1-3}
\begin{aligned}\sum_{i\in[4,15]}\mathcal{T}_{i\rightarrow h_{3}(3)}=2\times\left(|\mathcal{V}_{3}|-|\mathcal{V}_{2}|\right)+\sum_{j\in[1,3]}\mathcal{T}_{j\rightarrow h_{3}}(3).
\end{aligned}
\end{equation} 

Similarly, it is easy to verify that the next equation holds true.

\begin{equation}\label{eqa:MF-5-1-4}
\begin{aligned}\sum_{i\in[16,42]}\mathcal{T}_{i\rightarrow h_{4}(t)}&=2\times\left(|\mathcal{V}_{4}|-|\mathcal{V}_{2}|\right)+2\sum_{j\in[1,3]}\mathcal{T}_{j\rightarrow h_{4}}(4)\\
&\quad+\sum_{l\in[7,15]}\mathcal{T}_{l\rightarrow h_{4}}(4).
\end{aligned}
\end{equation}

Proceeding analogously, it is not difficult to obtain

\begin{equation}\label{eqa:MF-5-1-5}
\begin{aligned}\sum_{i\in[|\mathcal{V}_{t-1}|+1,|\mathcal{V}_{t}|]}\mathcal{T}_{i\rightarrow h_{t}}(t)&=2\left(|\mathcal{V}_{t}|-|\mathcal{V}_{t-1}|\right)\\
&+\sum_{j\in[2,t-1]}2^{t-1-j}\sum_{l\in[|\mathcal{V}_{j-1}|+1,|\mathcal{V}_{j}|]}\mathcal{T}_{l\rightarrow h_{t}}(t).
\end{aligned}
\end{equation}

With the help of previous results in \cite{Bollt-2005}, there is a scaling parameter $\alpha_{i}(t)$ such that $\mathcal{T}_{i\rightarrow h_{t+1}}(t+1)=\alpha_{i}(t+1)\mathcal{T}_{i\rightarrow h_{t}}(t)$ holds true when vertex $i$ is created before time step $t$. For instance, it is clear to the eye that $\mathcal{T}_{i\rightarrow h_{3}}(3)=\alpha_{i}(3)\mathcal{T}_{i\rightarrow h_{2}}(2)=\frac{25}{16}\mathcal{T}_{i\rightarrow h_{2}}(2)$ where $i=4,5,6$. Additionally, it follows from consequence in \cite{Bollt-2005} that scaling parameter $\alpha_{i}(t)$ is not only strictly smaller than $9/5$ but also larger than $1$ for $t>2$ and all vertices $i$ in question. Without loss of generality, we set $\alpha=\max_{t>2}\max_{i\in\mathcal{V}_{t-1}}\alpha_{i}(t)$. After that, we derive

\begin{equation}\label{eqa:MF-5-1-6}
\begin{aligned}\sum_{i\in[|\mathcal{V}_{t}|+1,|\mathcal{V}_{t+1}|]}&\mathcal{T}_{i\rightarrow h_{t+1}}(t+1)\\
&<3\alpha\sum_{i\in[|\mathcal{V}_{t-1}|+1,|\mathcal{V}_{t}|]}\mathcal{T}^{\ast}_{i\rightarrow h_{t}}(t)+(6-4\alpha)3^{t-1}.
\end{aligned}
\end{equation} 
and, further obtain 

\begin{equation}\label{eqa:MF-5-1-7}
\begin{aligned}\sum_{i\in[|\mathcal{V}_{t}|+1,|\mathcal{V}_{t+1}|]}&\mathcal{T}_{i\rightarrow h_{t+1}}(t+1)\\
&<6\times(3\alpha)^{t-1}+\frac{(6-4\alpha)(\alpha^{t-2}-1)3^{t-1}}{\alpha-1}.
\end{aligned}
\end{equation}

By definition, it is not hard to check the following inequality 

\begin{equation}\label{eqa:MF-5-1-8}
\begin{aligned}\sum_{i\in[|\mathcal{V}_{1}|+1,|\mathcal{V}_{t}|]}&\mathcal{T}_{i\rightarrow h_{t}}(t)
<\alpha\sum_{i\in[|\mathcal{V}_{1}|+1,|\mathcal{V}_{t-1}|]}\mathcal{T}_{i\rightarrow h_{t-1}}(t-1)\\
&+6\times(3\alpha)^{t-2}+\frac{(6-4\alpha)(\alpha^{t-3}-1)3^{t-2}}{\alpha-1}\\
&=6\alpha^{t-2}+\frac{3(3-2\alpha)\alpha^{t-3}(3^{t-2}-1)}{\alpha-1}\\
&+9\alpha^{t-2}(3^{t-2}-1)-\frac{3^{t-1}\left[1-\left(\frac{\alpha}{3}\right)^{t-2}\right]}{3-\alpha}.
\end{aligned}
\end{equation} 

So far, using Eqs.(\ref{eq:MF-2023-2-2}), (\ref{eqa:MF-5-0-1}) and (\ref{eqa:MF-5-1-8}), we obtain that in the limit $t\rightarrow\infty$, 

\begin{equation}\label{eqa:MF-5-1-9}
\langle\mathcal{T}_{t}\rangle<O(\alpha^{t})<O\left(|\mathcal{V}_{t}|^{2-\frac{\ln5}{\ln3}}\right)
\end{equation}
as desired. This completes the proof of Proposition 7. \qed 

In \cite{Zhang-2009}, the scale of average trapping time $\langle T_{t}\rangle$ for the typical Apollonian network $A_{t}$ is proven to follow $\langle T_{t}\rangle=O\left(|V_{t}|^{2-\frac{\ln5}{\ln3}}\right)$. So, the result above suggests that the proposed Type-II Apollonian network $\mathcal{A}_{t}$ has a faster transmit time than Type-I Apollonian network $A_{t}$. It is a little pit that we do not derive an exact solution of quantity $\langle\mathcal{T}_{t}\rangle$ for large parameter $t$. This is left for us and interested reader as the future work.

\subsection{Clique-based strategy}

In this subsection, we study another setting where the hub is still occupied by trap as studied above. The significant difference is that we bring in some relation between walkers. Concretely speaking, three walkers allocated on an arbitrary triangle, which contains three vertices created at the same time step, will together perform random walks and jump on a triangle. It is worth noting that three vertices in each anticipate triangle are also created at the same time step. In order to not cause ambiguity, we call this kind of trapping problem clique-based strategy (more precisely, $K_{3}$-based strategy where $K_{3}$ represents the complete graph of three vertices.), and, accordingly, make use of $\mathcal{T}^{\ast}_{i\rightarrow h_{t}}(t)$ to indicate the trapping time for walker starting out from vertex $i$ in network $\mathcal{A}_{t}$.       

\textbf{Proposition 8} The exact solution of average trapping time $\langle\mathcal{T}^{\ast}_{t}\rangle$ on network $\mathcal{A}_{t}$ is given by
\begin{equation}\label{eq:MF-5-2-0}
\langle\mathcal{T}^{\ast}_{t}\rangle=\frac{9^{t-1}+5\times3^{t-1}+9\left[\left(\frac{9}{5}\right)^{t-2}-\left(\frac{27}{5}\right)^{t-2}\right]}{4\times3^{t}-12}.
\end{equation}

\emph{Proof} As above, we first consider two concrete examples. In network $\mathcal{A}_{2}$, it is clear to see that 

\begin{equation}\label{eqa:MF-5-2-1}
\mathcal{T}^{\ast}_{i\rightarrow h_{2}}(2)=1,\quad i=4,5,6. 
\end{equation} 
Analogously, we consider network $\mathcal{A}_{2}$ and then derive 

\begin{equation}\label{eqa:MF-5-2-2}
\left\{\begin{aligned}&\mathcal{T}^{\ast}_{i\rightarrow h_{3}}(3)=\frac{9}{5},\quad i=4,5,6;\\
&\mathcal{T}^{\ast}_{i\rightarrow h_{3}}(3)=\frac{8}{5},\quad i=7,...,15.\\
\end{aligned}\right. 
\end{equation}  

In the meanwhile, we eatablish the next relationship

\begin{equation}\label{eqa:MF-5-2-3}
\begin{aligned}\sum_{i\in[4,15]}\mathcal{T}^{\ast}_{i\rightarrow h_{3}(3)}=|\mathcal{V}_{3}|-|\mathcal{V}_{2}|+\sum_{j\in[1,3]}\mathcal{T}^{\ast}_{j\rightarrow h_{3}}(3).
\end{aligned}
\end{equation} 

Along the same research line, it is easy to obtain

\begin{equation}\label{eqa:MF-5-2-4}
\begin{aligned}\sum_{i\in[|\mathcal{V}_{t-1}|+1,|\mathcal{V}_{t}|]}\mathcal{T}^{\ast}_{i\rightarrow h_{t}}(t)&=|\mathcal{V}_{t}|-|\mathcal{V}_{t-1}|\\
&+\sum_{j\in[2,t-1]}2^{t-1-j}\sum_{l\in[|\mathcal{V}_{j-1}|+1,|\mathcal{V}_{j}|]}\mathcal{T}^{\ast}_{l\rightarrow h_{t}}(t).
\end{aligned}
\end{equation} 

After that, we can easily build up a recurrence between quantities $\sum_{i\in[|\mathcal{V}_{t-1}|+1,|\mathcal{V}_{t}|]}\mathcal{T}^{\ast}_{i\rightarrow h_{t}}(t)$ and $\sum_{i\in[|\mathcal{V}_{t}|+1,|\mathcal{V}_{t+1}|]}\mathcal{T}^{\ast}_{i\rightarrow h_{t+1}}(t+1)$ as below 

\begin{equation}\label{eqa:MF-5-2-5}
\begin{aligned}\sum_{i\in[|\mathcal{V}_{t}|+1,|\mathcal{V}_{t+1}|]}&\mathcal{T}^{\ast}_{i\rightarrow h_{t+1}}(t+1)\\
&=\frac{27}{5}\sum_{i\in[|\mathcal{V}_{t-1}|+1,|\mathcal{V}_{t}|]}\mathcal{T}^{\ast}_{i\rightarrow h_{t}}(t)-\frac{3^{t}}{5}.
\end{aligned}
\end{equation} 
Notice that we have used a relation $\mathcal{T}^{\ast}_{i\rightarrow h_{t+1}}(t+1)=\frac{9}{5}\mathcal{T}^{\ast}_{i\rightarrow h_{t}}(t)$ where $i$ belongs to range $[4,|\mathcal{V}_{t}|]$, which is built based on consequence in \cite{Bollt-2005}. With initial condition $\mathcal{T}^{\ast}_{i\rightarrow h_{2}}(2)=1$ where $i\in[4,6]$, solving $\sum_{i\in[|\mathcal{V}_{t-1}|+1,|\mathcal{V}_{t}|]}\mathcal{T}^{\ast}_{i\rightarrow h_{t}}(t)$ from Eq.(\ref{eqa:MF-5-2-5}) yields

\begin{equation}\label{eqa:MF-5-2-6}
\sum_{i\in[|\mathcal{V}_{t-1}|+1,|\mathcal{V}_{t}|]}\mathcal{T}^{\ast}_{i\rightarrow h_{t}}(t)=\frac{9}{4}\left(\frac{27}{5}\right)^{t-2}+\frac{3^{t-1}}{4}.
\end{equation} 
To make further progress, we have 

\begin{equation}\label{eqa:MF-5-2-7}
\begin{aligned}\sum_{i\in[|\mathcal{V}_{1}|+1,|\mathcal{V}_{t}|]}\mathcal{T}^{\ast}_{i\rightarrow h_{t}}(t)&=\sum_{i\in[|\mathcal{V}_{1}|+1,|\mathcal{V}_{t-1}|]}\mathcal{T}^{\ast}_{i\rightarrow h_{t}}(t)\\
&\quad+\frac{9}{4}\left(\frac{27}{5}\right)^{t-2}+\frac{3^{t-1}}{4}\\
&=\frac{9}{5}\sum_{i\in[|\mathcal{V}_{1}|+1,|\mathcal{V}_{t-1}|]}\mathcal{T}^{\ast}_{i\rightarrow h_{t-1}}(t-1)\\
&\quad+\frac{9}{4}\left(\frac{27}{5}\right)^{t-2}+\frac{3^{t-1}}{4}.
\end{aligned}
\end{equation} 

Then, we derive the exact solution of quantity $\sum_{i\in[|\mathcal{V}_{1}|+1,|\mathcal{V}_{t}|]}\mathcal{T}^{\ast}_{i\rightarrow h_{t}}(t)$ as below  

\begin{equation}\label{eqa:MF-5-2-8}
\begin{aligned}\sum_{i\in[|\mathcal{V}_{1}|+1,|\mathcal{V}_{t}|]}\mathcal{T}^{\ast}_{i\rightarrow h_{t}}(t)&=\frac{9^{t-1}}{8}+\frac{5\times3^{t-1}}{8}\\
&+\frac{9}{8}\left[\left(\frac{9}{5}\right)^{t-2}-\left(\frac{27}{5}\right)^{t-2}\right].
\end{aligned}
\end{equation} 

By definition (\ref{eqa:MF-5-0-1}), we have 

\begin{equation}\label{eq:MF-5-2-9}
\begin{aligned}\langle\mathcal{T}^{\ast}_{t}\rangle&=\frac{\sum_{i\in[|\mathcal{V}_{1}|+1,|\mathcal{V}_{t}|]}\mathcal{T}^{\ast}_{i\rightarrow h_{t}}(t)}{|\mathcal{V}_{t}|-3}\\
&=\frac{\frac{9^{t-1}}{8}+\frac{5\times3^{t-1}}{8}+\frac{9}{8}\left[\left(\frac{9}{5}\right)^{t-2}-\left(\frac{27}{5}\right)^{t-2}\right]}{\frac{1}{2}\times3^{t}-\frac{3}{2}}
\end{aligned}
\end{equation}
Using some simple arithmetics, Eq.(\ref{eq:MF-5-2-9}) is simplified to lead to the same result as shown in Proposition 8. This is complete. \qed

From Eqs.(\ref{eq:MF-2023-2-2}) and (\ref{eq:MF-5-2-0}), one can see that for large parameter $t$, 

\begin{equation}\label{eq:MF-5-2-10}
\langle\mathcal{T}^{\ast}_{t}\rangle\sim 3^{t}=O(|\mathcal{V}_{t}|).
\end{equation}
This means that in this setting, average trapping time $\langle\mathcal{T}^{\ast}_{t}\rangle$ grows linearly with the number of vertices of network $\mathcal{A}_{t}$. 

Clearly, taking into account Eqs.(\ref{eq:MF-5-1-0}) and (\ref{eq:MF-5-2-10}), we firmly declare that in the trapping problem under consideration, the $K_{3}$-based strategy is more inferior than the vertex-based strategy on Type-II Apollonian network $\mathcal{A}_{t}$. More generally, we believe that the statement above also holds on some other networks while a mathematically rigorous proof is not provided here. The issue of how to verify this statement from the point of view of theory will be as our next move. In addition, other types of cliques are also selected as base to develop strategy when discussing trapping problem on networks. This is helpful to understand effect from the size of clique on average trapping time.

\section{Related work}

Nowadays, complex networks, as the powerful tool, have been used to model a wide variety of real-world complex systems, and succeeds in uncovering the structural characteristics and evolution mechanism on those systems \cite{Barab-2016}-\cite{Tyloo-2018}, such as, scale-free feature \cite{Albert-1999-1}, small-world property \cite{Watts-1998}, self-similarity \cite{Song-2005}, and so on. This triggers the increasing interest of the study of complex networks, and have drawn tremendous attention from various fields \cite{Barab-2016,Newman-2018,SouzaM-2023,Cheng-2020} including applied mathematics, statistical physics, computer science, etc. Roughly speaking, there are two mainstreams in the current study of complex networks \cite{Newman-2018}. The first is to focus on how to create theoretical models that can possess as many popularly-observed properties as possible. In other words, people who work in this first mainstream are concerned with topological structure of networks. In the other research direction, the study of various dynamics taking place on networks becomes the topic, including random walks \cite{Noh-2004}, trapping problem \cite{Zhang-2009}, and so forth.      

It is well known that among published theoretical models, BA-model attributed to Barabais and Albert \cite{Albert-1999-1} and WS-model established by Watts and Strogatz \cite{Watts-1998} are the most famous models. The former devotes to unveiling the mechanism behind scale-free networks that are ubiquitous in real world. The latter is used to describe the evolution of small-world networks. After that, a great number of theoretical models have been proposed, including Apollonian network \cite{Andrade-2005,Doye-2005}, Sierpinski gasket \cite{Chang-2007,Ri-2020}, Vicsek fractal \cite{Vicsek-1991,Ma-2022-TNES}, T-graph \cite{Agliari-2008,Ma-tkde-2020}, Dense network \cite{Ma-2020-T-1,Ma-pre-2020}, Triangle network \cite{Renou-2019} and so on. The previously-published results have shown that these models indeed capture some widely-observed structural properties of real-world networks, and have been successfully used to model some specific real-life network. For instance, Apollonian network is proven to have both scale-free feature and small-world property \cite{Andrade-2005,Doye-2005}, and can be utilized to depict the topological structure of brain \cite{Pellegrini-2007}. It should be mentioned that most of these models are planar. In fact, The family of planar graphs is a particularly fundamental yet significant family and is employed to model many important real-world graphs such as road networks, the layout of printed circuits, river networks upon the earth's surface, and so forth \cite{Mahapatra-2021,Dujmovic-2021,Rost-2020}. For example, T-graph has been utilized to describe the underlying structure of Peano river network \cite{Bartolo-2016}. Hence. it is of great interest to construct theoretical models having some structural properties observed in realistic networks.   

In the past decades, researchers have paid more attention on the study of a variety of dynamics occurring on complex networks \cite{Masuda-2017}. Among which, random walks on networks is the hot-topic in this field \cite{Noh-2004}, has been widely utilized in various science communities \cite{Yen-2010,Condamin-2008,Huang-2021}. As a specific case, trapping problem on complex networks has been widely studied in the literature \cite{Zhang-2009,Ma-2020-T-1}. Accordingly, previous researches show \cite{Masuda-2017} that for regular lattices, Sierpinski fractals and T-graph with number $|\mathcal{V}|$ of vertices, the associated average trapping times $\langle\mathcal{T}_{\mathcal{G}}\rangle$ all behave superlinearly with the number of vertices, i.e., $\langle\mathcal{T}_{\mathcal{G}}\rangle\sim |\mathcal{V}|^{\beta}$ where exponent $\beta$ is always greater than $1$. Besides that, the scale of average trapping time $\langle\mathcal{T}_{K_{|\mathcal{V}|}}\rangle$ for the complete graph $K_{|\mathcal{V}|}$ of $|\mathcal{V}|$ vertices has been proven to follow $\langle\mathcal{T}_{K_{|\mathcal{V}|}}\rangle\sim |\mathcal{V}|$. In theory, average trapping time is used as a metric to measure whether the underlying structure of network under consideration is or not beneficial to information diffusion \cite{Masuda-2017}. Given the number of vertices, the smaller average trapping time implies a better topological structure. Hence, the complete graph displays much better topological structure than other networks mentioned above. In \cite{Zhang-2009}, trapping problem on the typical Apollonian network is studied in more detail, and, consequently, average tapping time turns out to increases sublinearly with the number of vertices in which exponent is equal to $2-\frac{\ln5}{\ln2}$ in the limit of large graph size. So, the typical Apollonian network is better compared with the complete graph. Along the line of researches of such type, the problem of how to generate networks with smaller average trapping time is challenging and interesting from the theoretical viewpoint of information diffusion in a random walks based manner, and is worth making more efforts to deeply explore in the future.

\section{Conclusion}

To conclude, we present a theoretical model $\mathcal{A}_{t}$, called Type-II Apollonian network, based on the well-known Apollonian packing. We show that Type-II Apollonian network $\mathcal{A}_{t}$ is a maximal planar graph. Furthermore, network $\mathcal{A}_{t}$ turns out to be hamiltonian and eulerian. Next, we study some fundamental yet important structural parameters including average degree, degree distribution, clustering coefficient, diameter and Pearson correlation coefficient on network $\mathcal{A}_{t}$. The results demonstrate that the proposed network is sparse, has scale-free feature and small-world property, as well as displays disassortative mixing characteristic. Then, we consider the problem of enumeration of spanning trees of network $\mathcal{A}_{t}$, and derive the spanning trees entropy as well. The results show that Type-II Apollonian network is more reliable to a random removal of edges than the typical Apollonian network. Finally, we study trapping problem on network $\mathcal{A}_{t}$, and make use of average trapping time as a metric to verify that Type-II Apollonian network $\mathcal{A}_{t}$ has better structure for fast information diffusion than the typical Apollonian network.

\section*{Acknowledgment}

The research was supported by the Fundamental Research Funds for the Central Universities No. G2023KY05105, the National Key Research and Development Plan under grant 2020YFB1805400 and the National Natural Science Foundation of China under grant No. 62072010.

\section*{Supplementary Materials}

Here contains some useful materials omitted in main text. Specifically, we provide an illustration for proof of Lemma 2 and an illustrative example clarifying how to label vertices in network $\mathcal{A}_{t}$ using proper integer. 

We provide some additional materials in order to help reader to well understand the proof of Lemma 2. It should be mentioned that in the following figures (i.e., Figs. 9-12), we use a triangle to indicate a spanning tree of network $\mathcal{A}_{t}$, an edge along with an isolated vertex to represent a $2$-spanning-forest of network $\mathcal{A}_{t}$ in which one contains two greatest degree vertices (denoted by end-vertices of that edge) and the other includes the third, and three isolated vertices to stand for a $3$-spanning-forest of network $\mathcal{A}_{t}$ each of which only contains a greatest degree vertex. In Fig.13, we label each vertex in network $\mathcal{A}_{t}$ where $t=1,2,3$.

\begin{figure*}
\centering
\includegraphics[height=14cm]{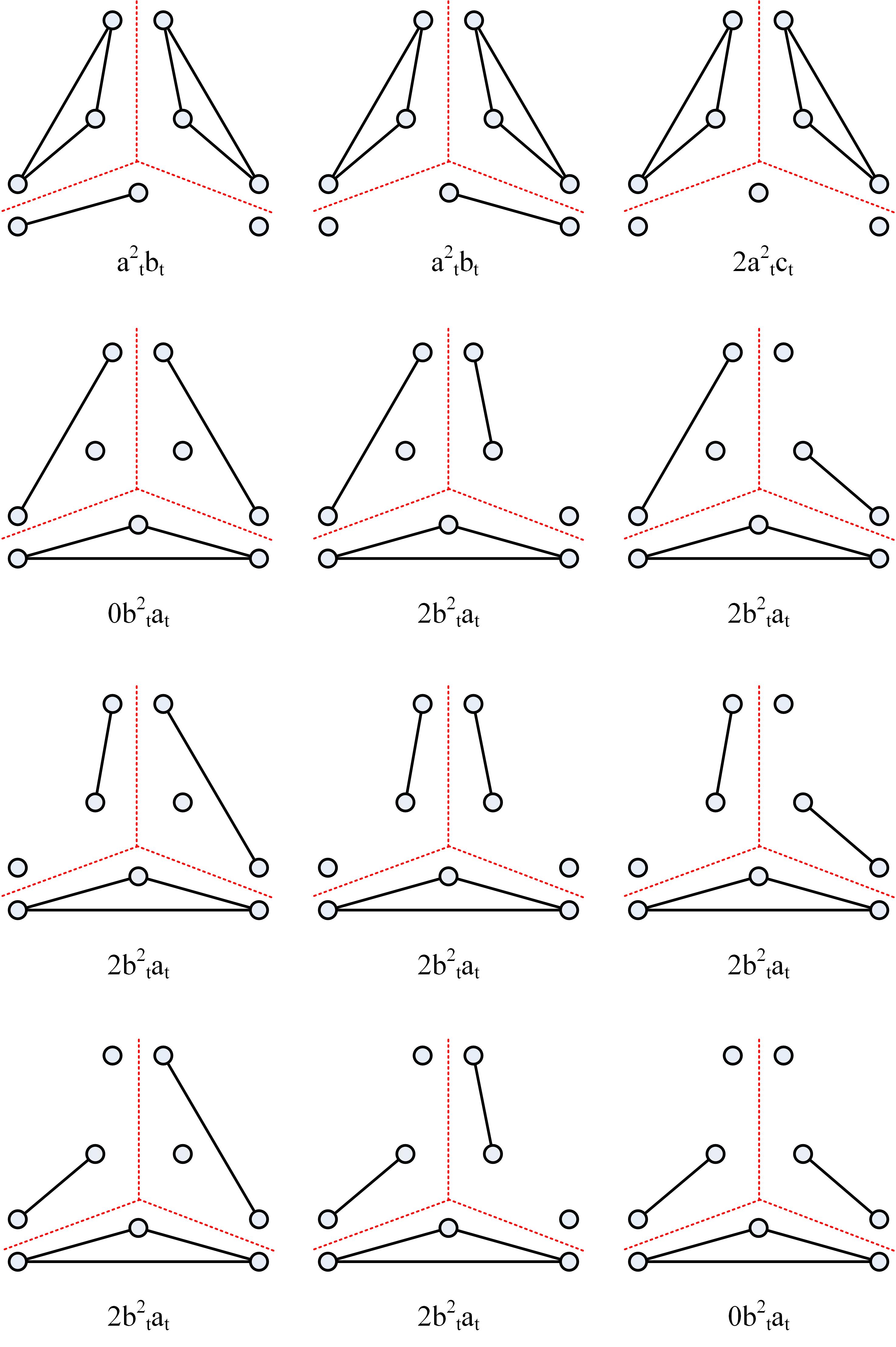}%\includegraphics[width=3in]{model2_ExpIncomeperTime_SymmetricVsAsymmetric.eps}
\caption{(Color online) The diagram of the contributions from quantities $s_{t}$ and $b_{t}$ to quantity $s_{t+1}$.}\label{fig:ExpIncomePerTime}
\end{figure*}

\begin{figure*}
\centering
\includegraphics[height=18cm]{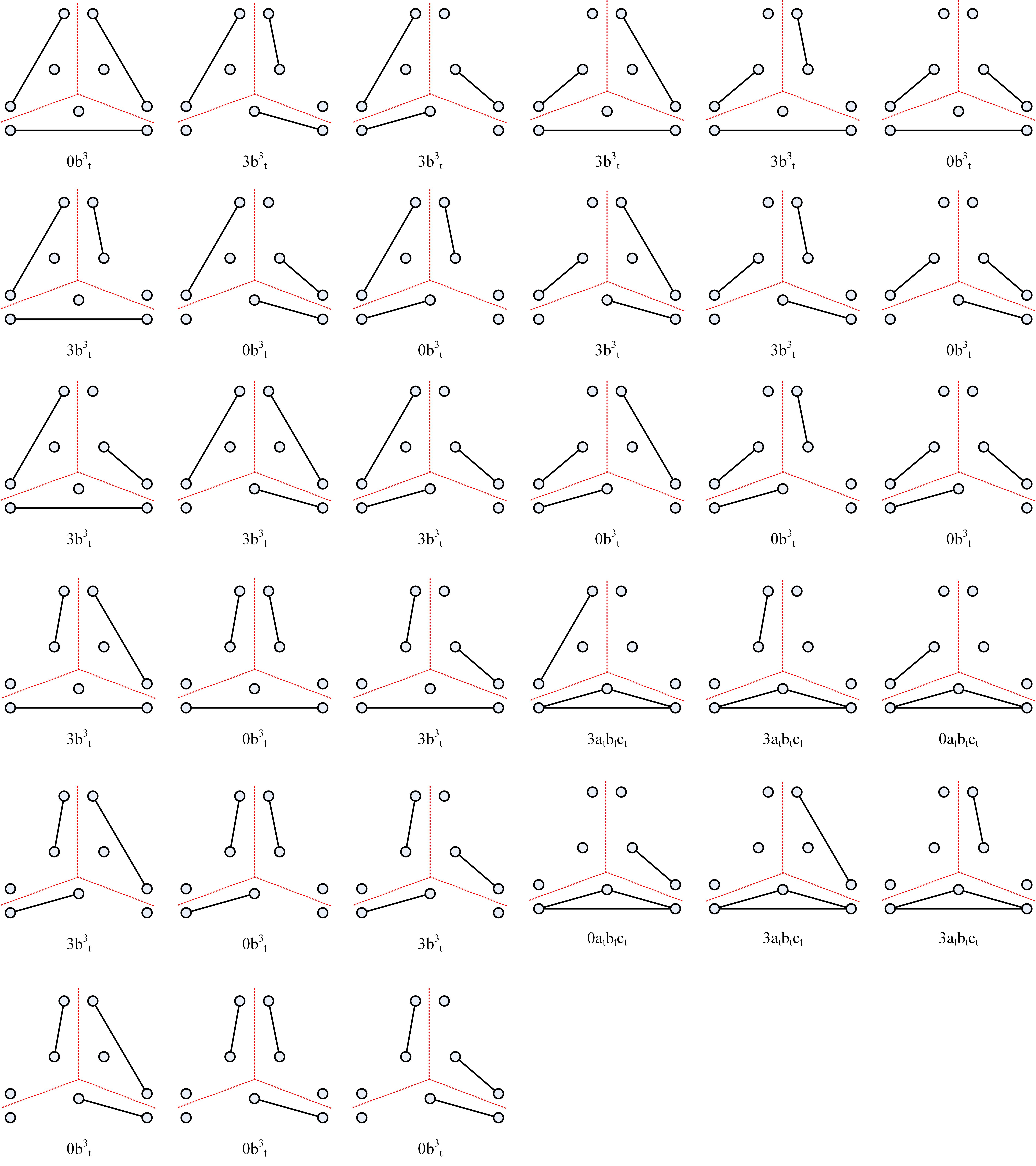}%\includegraphics[width=3in]{model2_ExpIncomeperTime_SymmetricVsAsymmetric.eps}
\caption{(Color online) The diagram of the contribution from quantities $s_{t}$, $b_{t}$ and $c_{t}$ to quantity $s_{t+1}$.}\label{fig:ExpIncomePerTime}
\end{figure*}

\begin{figure*}
\centering
\includegraphics[height=15cm]{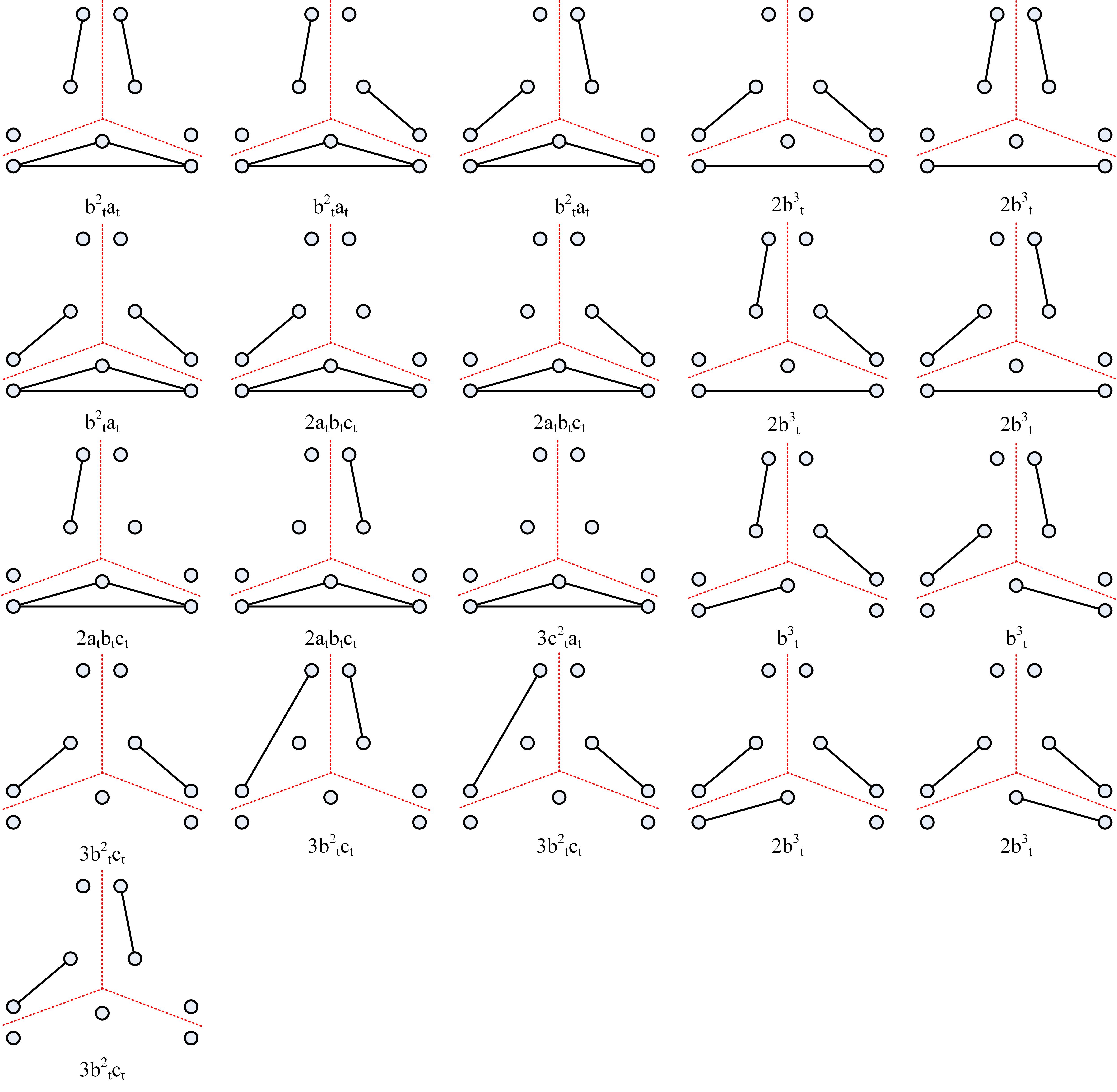}%\includegraphics[width=3in]{model2_ExpIncomeperTime_SymmetricVsAsymmetric.eps}
\caption{(Color online) The diagram of the contribution from quantities $s_{t}$, $b_{t}$ and $c_{t}$ to quantity $b_{t+1}$.}\label{fig:ExpIncomePerTime}
\end{figure*}

\begin{figure*}
\centering
\includegraphics[height=11cm]{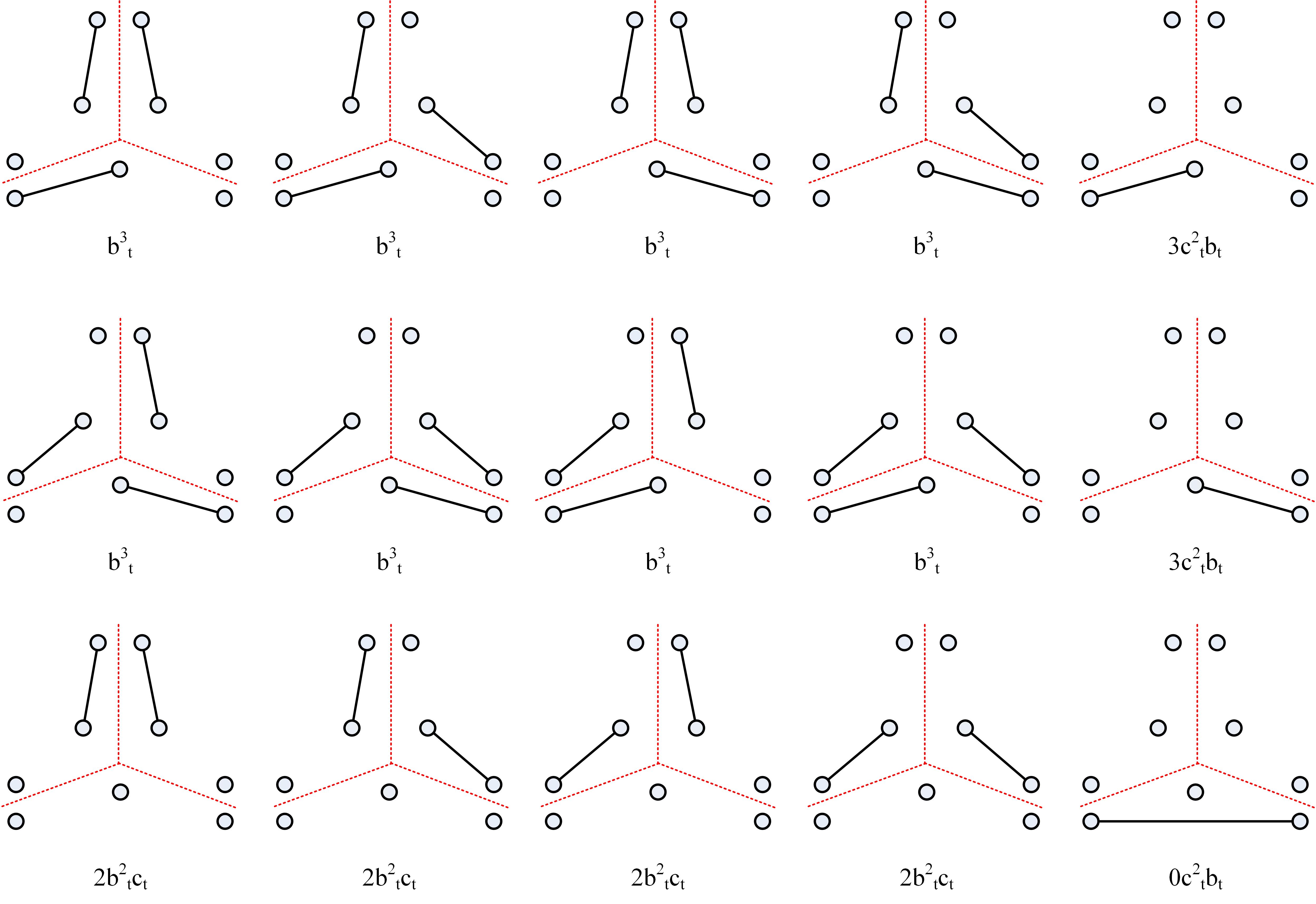}%\includegraphics[width=3in]{model2_ExpIncomeperTime_SymmetricVsAsymmetric.eps}
\caption{(Color online) The diagram of the contribution from quantities $b_{t}$ and $c_{t}$ to quantity $c_{t+1}$.}\label{fig:ExpIncomePerTime}
\end{figure*}

\begin{figure*}
\centering
\includegraphics[height=13cm]{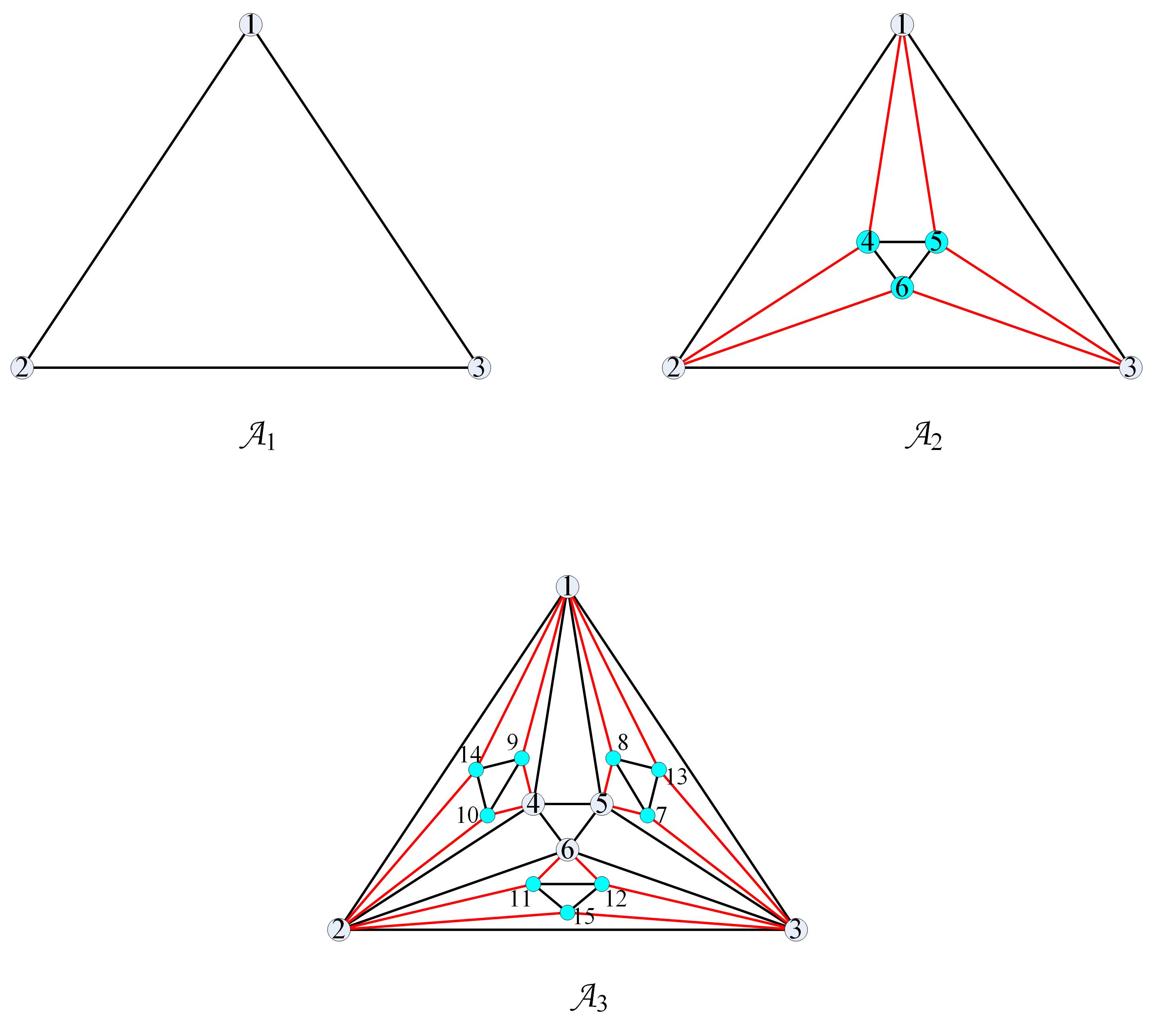}%\includegraphics[width=3in]{model2_ExpIncomeperTime_SymmetricVsAsymmetric.eps}
\caption{(Color online) The diagram of how to label each vertex in network $\mathcal{A}_{t}$ where $t=1,2,3$.}\label{fig:ExpIncomePerTime}
\end{figure*}


{\footnotesize
\begin{thebibliography}{9}

\setlength{\parskip}{0pt}


\bibitem{Barab-2016} A.-L. Barab\'{a}si. Network Science. Cambridge University Press (2016).

\bibitem{Newman-2018} M. E. J. Newman. Networks. Oxford University Press (2018).

\bibitem{Zou-2019} Y. Zou, R. V. Donner, N. Marwan, J. F. Donges, and J. Kurths. Complex network approaches to nonlinear time series analysis. Physics Reports, 2019, 787: 1-97.

\bibitem{Lynn-2020} C. W. Lynn, L. Papadopoulos, A. E. Kahn, and D. S. Bassett. Human information processing in complex networks. Nature Physics, 2020, 16(9): 965-973.

\bibitem{Tyloo-2018} M. Tyloo, T. Coletta, and P. Jacquod. Robustness of synchrony in complex networks and generalized Kirchhoff indices. Physical Review Letters, 2018, 120(8): 084101.

\bibitem{Albert-1999-1} A.-L. Barab\'{a}si, R. Albert. Emergence of scaling in random networks. Science, 1999, 5439: 509-512.

\bibitem{Watts-1998} D. J. Watts, S. H. Strogatz. Collective dynamics of `small-world' networks. Nature, 1998, 393: 440-442. 

\bibitem{Song-2005} C. Song, S. Havlin, and H. A. Makse. Self-similarity of complex networks. Nature, 2005, 433(7024): 392-395.

\bibitem{Amaral-2000} L. A. N. Amaral, A. Scala, M. Barthelemy, and H. E. Stanley. Classes of small-world networks. Proceedings of the National Academy of Sciences, 2000, 97(21): 11149-11152.

\bibitem{Latora-2001} V. Latora, M. Marchiori. Efficient behavior of small-world networks. Physical Review Letters, 2001, 87(19): 198701.

\bibitem{Andrade-2005} J. S. Andrade Jr, H. J. Herrmann, R. F. S. Andrade, and L. R. da Silva. Apollonian networks: simultaneously scale-free, small world, Euclidean, space filling, and with matching graphs. Physical Review Letters, 2005, 94: 018702.

\bibitem{Doye-2005} J. P. K. Doye, C. P. Massen. Self-similar disk packings as model spatial scale-free networks. Physical Review E, 71, (2005) 016128.

\bibitem{Chang-2007} S. C. Chang, L. C. Chen, and W. S. Yang. Spanning trees on the Sierpinski gasket. Journal of Statistical Physics, 2007, 126: 649-667. 

\bibitem{Ri-2020} S. I. Ri. Fractal functions on the Sierpinski Gasket. Chaos, Solitons and Fractals, 2020, 138: 110142. 

\bibitem{Vicsek-1991} A.-L Barab\'{a}si, T. Vicsek. Multifractality of self-affine fractals. Physical Review A, 1991, 44(4): 2730.

\bibitem{Ma-epl-2021} F. Ma, X. Wang, P. Wang, and X. Luo. Random walks on the generalized Vicsek fractal. Europhysics Letters, 2021, 133(4): 40004.

\bibitem{Agliari-2008} E. Agliari. Exact mean first-passage time on the T-graph. Physical Review E, 2008, 77(1): 011128.

\bibitem{Ma-tkde-2020} F. Ma, P. Wang, and X. Luo. A method for geodesic distance on subdivision of trees with arbitrary orders and their applications. IEEE Transactions on Knowledge and Data Engineering, 2022, 34(5): 2063-2075. 

\bibitem{Ma-2020-T-1} F. Ma, P. Wang. Determining exact solutions for structural parameters on hierarchical networks with density feature. The Computer Journal, 2021, 64(9): 1412-1424.

\bibitem{Ma-pre-2020} F. Ma, X. Wang, P. Wang, and X. Luo. Dense networks with scale-free feature. Physical Review E, 2020, 101(5): 052317.

\bibitem{Liu-2020} J. B. Liu, Z. Raza, and M. Javaid. Zagreb connection numbers for cellular neural networks. Discrete Dynamics in Nature and Society, 2020, 2020: 1-8.

\bibitem{Mahapatra-2021} T. Mahapatra, S. Sahoo, G. Ghorai, and M Pal. Interval valued m-polar fuzzy planar graph and its application. Artificial Intelligence Review, 2021, 54: 1649-1675.

\bibitem{Dujmovic-2021} V. Dujmovic, L. Esperet, C. Gavoille, J. Gwenael, P. Micek, and P. Morin. Adjacency labelling for planar graphs (and beyond). Journal of the ACM (JACM), 2021, 68(6): 1-33.

\bibitem{Newman-2011} M. E. J. Newman, A.-L. Barab\'{a}si, and D. J. Watts. The structure and dynamics of networks. Princeton University Press, 2011.

\bibitem{Masuda-2017} N. Masuda, M. A. Porter, and R. Lambiotte. Random walks and diffusion on networks. Physics reports, 2017, 716: 1-58.

\bibitem{Yen-2010} L. Yen, M. Saerens, and F. Fouss. A link analysis extension of correspondence analysis for mining relational databases. IEEE Transactions on Knowledge and Data Engineering, 2010, 23(4): 481-495.

\bibitem{Condamin-2008} S. Condamin, V. Tejedor, R. Voituriez, O. B\'{e}nichou, and J. Klafter. Probing microscopic origins of confined subdiffusion by first-passage observables. Proceedings of the National Academy of Sciences, 2008, 105(15): 5675-5680.

\bibitem{Huang-2021} Z. Huang, A. Silva, and A. Singh. A broader picture of random-walk based graph embedding. Proceedings of the 27th ACM SIGKDD conference on knowledge discovery data mining. 2021: 685-695.

\bibitem{Loverdo-2008} C. Loverdo, O. B\'{e}nichou, M. Moreau, and R. Voituriez. Enhanced reaction kinetics in biological cells. Nature Physics, 2008, 4(2): 134-137.

\bibitem{Zhao-2022} J. Zhao, T. Wen, H. Jahanshahi, and K. H. Cheong. The random walk-based gravity model to identify influential nodes in complex networks. Information Sciences, 2022, 609: 1706-1720.

\bibitem{Zaheer-2022} M. Zaheer, K. Marino, W. Grathwohl, J. Schultz, W. Shang, S. Babayan, A.Ahuja, I. Dasgupta, C. Kaeser-Chen, and R. Fergus. Learning to Navigate Wikipedia by Taking Random Walks. Advances in Neural Information Processing Systems, 2022, 35: 1529-1541.

\bibitem{Kasner-1943} E. Kasner, F. D. Supnick. The Apollonian packing of circles. Proceedings of the National Academy of Sciences, 1943, 29(11): 378-384.

\bibitem{Bondy-2008} J. A. Bondy. U. S. R. Murty. Graph Theory. Springer (2008).

\bibitem{Fu-2005} J. S. Fu. Hamiltonicity of the WK-recursive network with and without faulty nodes. IEEE Transactions on Parallel and Distributed Systems, 2005, 16(9): 853-865.


\bibitem{Newman-2003} M. E. J. Newman. The structure and function of complex networks. SIAM Review, 2003, 45(2): 167-256.

\bibitem{Grabow-2015} C. Grabow, S. Grosskinsky, J. Kurths, and M. Timme. Collective relaxation dynamics of small-world networks. Physical Review E, 2015, 91(5): 052815.

\bibitem{Newman-2002} M. E. J. Newman. Assortative mixing in networks. Physical Review Letters, 2002, 89(20): 208701.

\bibitem{Ma-TCS-2018} F. Ma, B. Yao. An iteration method for computing the total number of spanning trees and its applications in graph theory. Theoretical Computer Science, 2018, 708: 46-57.

\bibitem{Neumann-2007} F. Neumann, I. Wegener. Randomized local search, evolutionary algorithms, and the minimum spanning tree problem. Theoretical Computer Science, 2007, 378(1): 32-40.

\bibitem{Zhang-2014} Z. Zhang, B. Wu, and F. Comellas. The number of spanning trees in Apollonian networks. Discrete Applied Mathematics, 2014, 169: 206-213.

\bibitem{Zhang-2013}  J. Zhang, W. Sun, and G. Xu. Enumeration of spanning trees on Apollonian networks. Journal of Statistical Mechanics: Theory and Experiment, 2013, 2013(09): P09015.

\bibitem{Li-2021} X. Y. Li, W. Lin, X. Liu, C.-K. Lin, K.-J. Pai, and J.-M. Chang. Completely independent spanning trees on BCCC data center networks with an application to fault-tolerant routing. IEEE Transactions on Parallel and Distributed Systems, 2021, 33(8): 1939-1952.

\bibitem{G-J-S-2003} G. J. Szab\'{o}, M. Alava, and J. Kert\'{e}sz. Geometry of minimum spanning trees on scale-free networks. Physica A: Statistical Mechanics and its Applications, 2003, 330: 31-36.

\bibitem{N-T-A-E-2006} N. Takashi, A. E. Motter. Synchronization is optimal in nondiagonalizable networks. Physical Review E, 2006, 73: 065106.

\bibitem{Afshari-2020} A. Afshari, M. Karrari, H. R. Baghaee, and G. B. Gharehpetian. Distributed fault-tolerant voltage/frequency synchronization in autonomous AC microgrids. IEEE Transactions on Power Systems, 2020, 35(5): 3774-3789.

\bibitem{P-M-2000} P. Marchal. Loop-erased random walks, spanning trees and Hamiltonian cycles. Electronic Communications in Probability, 2020, 5: 39-50.

\bibitem{Alev-2020} V. L. Alev, L. C. Lau. Improved analysis of higher order random walks and applications. Proceedings of the 52nd Annual ACM SIGACT Symposium on Theory of Computing. (2020) 1198-1211.

\bibitem{Lyons-2005} R. Lyons. Asymptotic enumeration of spanning trees. Combinatorics, Probability and Computing, 2005, 14(4): 491-522.

\bibitem{Noh-2004} J. D. Noh, H. Rieger. Random walks on complex networks. Physical Review Letters, 2004, 92(11): 118701.

\bibitem{Montroll-1969} E. W. Montroll. Random walks on lattices. III. Calculation of first-passage times with application to exciton trapping on photosynthetic units. Journal of Mathematical Physics, 1969, 10(4): 753-765.

\bibitem{Haggstrom-2002} O. Haggstrom. Finite Markov chains and algorithmic applications. Cambridge University Press, 2002.

\bibitem{Bollt-2005} E. M. Bollt, D. ben-Avraham. What is special about diffusion on scale-free nets?. New Journal of Physics, 2005, 7(1): 26.

\bibitem{Zhang-2009} Z. Zhang, J. Guan, W. Xie, Y. Qi, and S. Zhou. Random walks on the Apollonian network with a single trap. Europhysics Letters, 2009, 86(1): 10006.

\bibitem{SouzaM-2023} R. M. D'Souza, M. di Bernardo, and Y. Y. Liu. Controlling complex networks with complex nodes. Nature Reviews Physics, 2023, 5(4): 250-262.

\bibitem{Cheng-2020} F. Cheng, C. Wang, X. Zhang, and Y. Yang. A local-neighborhood information based overlapping community detection algorithm for large-scale complex networks. IEEE/ACM Transactions on Networking, 2020, 29(2): 543-556.

\bibitem{Ma-2022-TNES} F. Ma, P. Wang, and X. Luo. Random walks on stochastic generalized vicsek fractalnetworks: Analytic solution and simulations. IEEE Transactions on Network Science and Engineering, 2022, 9(3): 1335-1345.

\bibitem{Renou-2019} M.-O. Renou, E. Baumer, S. Boreiri, N. Brunner, N. Gisin, and S. Beig. Genuine quantum nonlocality in the triangle network. Physical Review Letters, 2019, 123(14): 140401.

\bibitem{Pellegrini-2007} G. L. Pellegrini, L. de Arcangelis, H. J. Herrmann, and C. Perrone-Capano. Modelling the brain as an Apollonian network. arXiv preprint q-bio/0701045, 2007.

\bibitem{Rost-2020} M. Rost, S. Schmid. On the hardness and inapproximability of virtual network embeddings. IEEE/ACM Transactions on Networking, 2020, 28(2): 791-803.

\bibitem{Bartolo-2016} S.D. Bartolo, F. Dell\'{a}ccio, G. Frandina, G. Moretti, S. Orlandini, and M. Veltri. Relation between grid, channel, and Peano networks in high-resolution digital elevation models. Water Resources Research,  2016, 52(5): 3527-3546.



\end{thebibliography}
}
\end{document}